\documentclass[aps,prl,twocolumn,superscriptaddress,
groupedaddress,10pt]{revtex4}
\usepackage{amsmath}
\usepackage{amssymb}
\usepackage{graphicx}
\usepackage{epsfig}
\usepackage{dcolumn}
\usepackage{bm}
\usepackage{bm}
\usepackage{blindtext}
\usepackage{natbib}
\usepackage{siunitx}

\usepackage{multirow}
\usepackage{bbm}

\usepackage[usenames,dvipsnames]{xcolor}
\usepackage{amsthm}
\usepackage{booktabs}
\usepackage{hyperref}
\usepackage{tikz}
\usetikzlibrary{calc}
\usepackage[printwatermark]{xwatermark}
\usepackage{mathtools}

\def\be{\begin{equation}} \def\ee{\end{equation}}
\def\bea{\begin{eqnarray}} \def\eea{\end{eqnarray}}

\def\nn{\nonumber}

\newcommand{\ket}[1]{| #1 \rangle}
\newcommand{\bra}[1]{\langle #1 |}

\begin{document}

\title{Single branch of chiral Majorana modes from doubly degenerate Fermi surfaces}
\author{Wang Yang}
\affiliation{Department of Physics, University of California,
San Diego, California 92093, USA}
\author{Chao Xu}
\affiliation{Department of Physics, University of California,
San Diego, California 92093, USA}
\author{Congjun Wu}
\affiliation{Department of Physics, University of California,
San Diego, California 92093, USA}

\begin{abstract}
Majorana fermions are often proposed to be realized by first singling out
one Fermi surface without spin degeneracy via spin-orbit coupling, 
 and then imposing boundaries or defects.
In this work, we take a different route starting with two degenerate Fermi surfaces
without spin-orbit coupling,
and show that by the method of ``kink on boundary",  the dispersive chiral Majorana fermions can be realized in superconducting systems with $p\pm is$ pairings.
The surfaces of these systems develop spontaneous magnetizations whose
directions are determined by the boundary orientations and the phase difference
between the $p$ and $s$-component gap functions.
Along the magnetic domain walls on the surface, there exist chiral Majorana
fermions propagating unidirectionally, which can be conveniently dragged and
controlled by external magnetic fields.
Furthermore, the surface magnetization is shown to be a magnetoelectric effect
based on a Ginzburg-Landau free energy analysis.
We also discuss how to use the proximity effects to realize chiral Majorana
fermions by performing the ``kink on boundary" method.
\end{abstract}
\maketitle

Majorana fermions are their own anti-particles which were first introduced
to high energy physics  \cite{Majorana1937}.
In the past decade, they have been intensively investigated in the context
of condensed matter physics \cite{Nayak2008}.
The braiding of Majorana particles exhibits non-Abelian statistics
\cite{MooreRead1991,Nayak1996,Ivanov2001,Stern2004},
which is distinct from the usual Fermi and Bose statistics
and can be applied for quantum
computations \cite{Kitaev2003,Kitaev2006,Stone2006,Alicea2011,Halperin2012}.
The topological nature of Majorana modes makes the brading and
fusion operations robust from the decoherence processes which are detrimental
to the realization of quantum computers.

The Majorana modes are proposed to exist in the $\nu=\frac{5}{2}$
fractional quantum Hall state \cite{MooreRead1991}.
There have also been considerable interests in studying Majorana fermions
in topological superconducting systems \cite{Hasan2010,Qi2011,Ando2015,Sato2016,Chiu2016,Sato2017}.
Majorana fermions appear on boundaries, in vortex cores, and
at defects of topological superconducting systems
\cite{Read2000,Kitaev2001,Ivanov2001,DasSarma2006,Fu2008,Sau2010,Teo2010}.
Chiral Majorana fermion has been proposed to emerge in the quantum
anomalous Hall insulator
in proximity with an $s$-wave superconductor\cite{Yu2010,Qi2010,Chung2011}.
The interaction effects in Majorana fermions have also been discussed
by various authors \cite{Cheng2009,Li2013,Potter2014}.
Recent experiments have provided evidence to the existence of Majorana
zero modes and chiral Majorana fermions in condensed matter
systems \cite{Doh2005,Sun2016,Jack2019}.

A Majorana fermion is half of a usual fermion in view of the degrees
of freedom that it contains.
Since electrons have two spin degrees of freedom, the chiral Majorana
fermions can only be obtained by a ``half of half" method.
Typically, the first ``half" is achieved by singling out a non-degenerate
Fermi surface in spin-orbit coupled systems, which becomes effectively
single-component.
The second ``half" is performed by imposing boundaries or defects
to generate zero modes.

On the other hand, there have been considerable experimental and theoretical interest in studying
superconducting states with competing singlet and triplet pairings \cite{Sasaki2001,Wu2010,Fu2010,Sasaki2011,Kobayashi2011,Kirzhner2012,Sasaki2012,Novak2013,Hinojosa2014,Zhou2015,Wang2017}.
A spontaneously time-reversal symmetry breaking mixing is energetically favored
 exhibiting $\pm \frac{\pi}{2}$ phase difference between
the gap functions in these two different channels \cite{Wu2010}.
This class of novel pairing states have been proposed in the ultra-cold electric dipolar fermion systems \cite{Wu2010},
in cold fermion systems under the $p$-wave Feshbach resonances \cite{Zhou2015},
in the iron-pnicitide superconductors \cite{Hinojosa2014},
and in the inversion symmetry breaking superconducting
systems \cite{Wang2017}.
Such kind of time reversal symmetry breaking superconductors host gapped Dirac cones on the surface with nontrivial gravitational responses and thermal Hall effects \cite{Ryu2012,Qi2013,Shiozaki2014,Goswami2014,Goswami2015,Stone2016,Wang2017}.
Recently, there has been strong experimental evidence for unconventional and time-reversal symmetry breaking pairing
in the superconducting state of the noncentrosymmetric material Re$_{0.82}$Nb$_{0.18}$
\cite{Shang2018},
which is very likely of a mixed singlet and triplet nature \cite{Sundar2019}.

In this article, we analyze the formation of chiral Majorana fermions
in superconductors with mixed singlet and triplet pairings
of the $p\pm is$ type.
Different from previous works, there are two degenerate Fermi surfaces
without any spin-orbit coupling,
and the strategy of ``half of half" is implemented as ``kink on boundary".
The boundaries of the $p\pm is$ superconductors are shown to be spontaneously spin polarized, and the magnetizations are opposite for $p+is$ and $p-is$ pairings.
As a result, the ``kink" formed by the domain wall between the $p+ is$ and $p-is$ superconducting regions on the surface
is also a magnetization domain wall.
We show that there exists a chiral Majorana fermion propagating unidirectionally along the ``kink".
The spirit of such ``kink on boundary" method is similar to realizing Majorana corner and hinge modes
in the high order topological systems \cite{Benalcazar2017,Wang2018,Zhu2019,Pan2019,Volpez2019,Bultinck2019,Wu2019,Zhang2019}.
However, the chiral Majorana fermions realized using our method are mobile in the sense that they can be dragged and controlled by external magnetic fields,
hence convenient for braiding purposes.
In addition, we show that the spontaneous surface magnetization is a manifestation
of a novel magnetoelectric effect based on a Ginzburg-Landau free energy analysis.
Finally, we also discuss how to use the proximity effects in triplet superconductors including Cu$_x$Bi$_2$Se$_3$ and Sn$_{1-x}$In$_x$Te to realize chiral Majorana fermions by
performing the ``kink on boundary" method.

We consider a gap function structure with a dominate $p$-wave
component mixed with a $s$-wave one.
It typically prefers the time-reversal symmetry breaking pairing pattern
$p\pm is$ \cite{Wu2010}.
The corresponding gap function matrix reads $\Delta_{\alpha\beta}(\vec k)=
\Delta_s (\vec k) (i\sigma_2)_{\alpha\beta}+\Delta_{p} (\vec k)\hat d (\vec k)
\cdot (i\vec \sigma \sigma_2)_{\alpha\beta}$
where  $\hat d(\vec k)$ is a unit real vector and
$\vec \sigma$'s are the Pauli matrices in spin space.
Only when the phase difference between $\Delta_s$ and $\Delta_p$
equals to $\pm \frac{\pi}{2}$, $\Delta_{\alpha\beta}$ is
proportional to a unitary matrix.
Typically, unitary pairings are energetically more favorable over
non-unitary ones \cite{Leggett1975}.
To see this, consider the following Ginzburg-Landau free energy
\bea
F&=& -\alpha_s \Delta_s^{*}\Delta_s-\alpha_p \Delta_p^{*} \Delta_p+\beta_s |\Delta_s|^4+\beta_p |\Delta_p|^4\nn\\
&&+\gamma_1 |\Delta_p|^2  |\Delta_s|^2 +\gamma_2 (\Delta_p^{*} \Delta_p^{*} \Delta_s \Delta_s+c.c),
\label{eq:GL}
\eea
in which both $\alpha_s$ and $\alpha_p$ are negative, signaling instabilities in both $s$- and $p$-wave channels.
Eq. (\ref{eq:GL}) is the most general form of the free energy up to quartic order which respects time reversal and inversion symmetries.
Since $\gamma_2$ is positive  generically \cite{Wang2017}, a $\pm \pi/2$ phase difference between $\Delta_s$ and $\Delta_p$ is energetically favored.
Up to an overall gauge transformation, the $p$- and $s$-wave components can be fixed as real and imaginary, respectively.
Both time reversal ($T$) and inversion ($P$) symmetries are spontaneously
broken in the $p\pm is$ pairing states.
Nevertheless, the system is invariant up to an overall phase
under the $PT$-transformation, {\it i.e.}, the combined parity and
time-reversal operations.

For simplicity, we start with a $1D$ $p_z\pm is$ superconductor, whose Bogoliubov-de Gennes (BdG) Hamiltonian reads
\begin{eqnarray}
H_{1d}&=&\frac{1}{2}\int dz \psi^{\dagger}(z)\Big( \big(-\frac{\hbar^2}{2m} \partial_z^2-\mu(z)\big) \tau_3-\Delta_s\sigma_2\tau_1\nn\\
&&-\frac{\Delta_p}{k_f} i\partial_z\sigma_1\tau_1\Big)\psi(z),
\label{eq:1DHam}
\end{eqnarray}
in which  $\psi(z)=(c^{\dagger}_{\uparrow}(z)
\,c^{\dagger}_{\downarrow}(z)\,c_{\uparrow}(z)\,c_{\downarrow}(z))^T$;
$\tau_i$'s are the Pauli matrices in the Nambu space;
$k_f$ is the Fermi wavevector;
$\Delta_s$ and $\Delta_p$ represent the singlet and triplet pairing
gap functions which are assumed real without loss of generality.
The BdG Hamiltonian Eq. (\ref{eq:1DHam}) possesses
the particle-hole symmetry $P_h H_{1d} P_h^{-1}=-H_{1d}$
where $P_h$ is an anti-unitary transformation defined as
$P_h\psi^{\dagger}(z)P_h^{-1}=\psi^{\dagger}(z) \sigma_0 \tau_1 K$,
with $K$ the complex conjugate operation.
The triplet pairing pattern in Eq. (\ref{eq:1DHam}) corresponds
to the $d$-vector configuration $\hat d\parallel \hat z$, hence
the $z$-component of spin is conserved represented
as $S_z=\frac{1}{4}\sigma_3 \tau_3$.
In the absence of $\Delta_s$, the system preserves time-reversal
symmetry $T\psi^{\dagger}(z)T^{-1}=\psi^{\dagger}(z)i\sigma_2 \tau_0 K$,
and there exists a chiral operator $C_{ch}=-iTP_h=\sigma_2\tau_1$ anti-commuting with the Hamiltonian.
The chiral operator in general maps positive energy states to negative energy states,
but becomes a symmetry for the zero modes.
In what follows, an open boundary condition is imposed along the $z$-direction
at the upper ($z=\frac{L}{2}$) and lower ($z=-\frac{L}{2}$) edges,
with $\mu(z)=\frac{\hbar^2}{2m} k_f^2$ at $|z|<\frac{L}{2}$
and $\mu(z)=-\infty$ at $|z|>\frac{L}{2}$, where $L$ is the system size.

When $\Delta_s=0$, there exist two Majorana zero modes at each edge
of the system.
The associated creation operators for the four Majorana modes are $\gamma^{a,\dagger}_{\lambda}=\int dz\psi^{\dagger}(z)\Psi^{a}_{\lambda}(z)$,
in which $a=+(-)$ for upper (lower) edge and
$\lambda=\pm$ labelling the two zero modes at each edge.
The zero mode wavefunctions $\Psi^{a}_{\lambda}$ are
solved as $\Psi^{a}_{+}(z)=\frac{1}{\sqrt{2}}(e^{-ia\frac{\pi}{4}},
0,0,e^{ia\frac{\pi}{4}})^T u_{+}(z)$
and
$\Psi^{a}_{-}(z)=\frac{1}{\sqrt{2}}(0,e^{-ia\frac{\pi}{4}},
e^{ia\frac{\pi}{4}},0)^T u_{-}(z)$, respectively,
where $u_{a}(z)$ is the envelope function with the expression
given in Supplementary Materials (SM) Sect. I \cite{suppl}.
Since $[C_{ch},S_z]=0$, the wavefunctions of the four zero modes can
be chosen as the simultaneous eigenstates of $C_{ch}$ and $S_z$:
\begin{eqnarray}
C_{ch} \Psi^{a}_{\lambda}=a\lambda \Psi^{a}_{\lambda}, &
S_{z} \Psi^{a}_{\lambda}=\frac{1}{2}\lambda \Psi^{a}_{\lambda}.
\label{eq:goodquantum1}
\end{eqnarray}
Furthermore, there exists an emergent supersymmetry expressed as $\gamma^{a}_{+}=\gamma^{a,\dagger}_{-}$\cite{Qi2009}.

When $\Delta_s\neq 0$, the four modes become gapped,
and a spontaneous magnetization develops on the edge.
Since the singlet pairing component in $H_{1d}$ is $-\Delta_sC_{ch}$,
 $C_{ch}$ and $S_z$ still form a complete set of good quantum numbers for the four modes as expressed in Eq. (\ref{eq:goodquantum1}).
Without loss of generality, let us consider the case of $\Delta_s>0$.
$\gamma^{a,\dagger}_{\lambda=a}$ represents quasiparticle annihilation operator, since $[H_1,\gamma^{a,\dagger}_{\lambda=a}]=-\Delta_s \gamma^{a,\dagger}_{\lambda=a}$.
The projection of the $S_z$ operator to the edge state subspace can be expressed as $S_z=-\frac{1}{2}
a\Big(\gamma^{a,\dagger}_{\lambda=-a}\gamma^{a,\dagger}_{\lambda=a}
-\frac{1}{2}\Big)$,
hence, 
\bea
\bra{G}S_z\ket{G}=\frac{1}{4}a,
\eea
 where $\ket{G}$ is the ground state of the system.
 This result shows the fractionalization of the $S_z$ eigenvalue to $\pm \frac{1}{4}$ on the boundary.
Therefore, the upper and lower edges carry spontaneous magnetizations along the $z$-direction,
and they are oppositely magnetized as enforced by the $PT$-symmetry.
Define the $PT$-operation as $\mathcal{S}=GPT$,
where $G:c_{\sigma}(z)\rightarrow ic_{\sigma}(z)$ ($\sigma=\uparrow,\downarrow$) is a gauge transformation.
The operation $\mathcal{S}$ flips both $a$ and $\lambda$ and
maintains the $C_{ch}$ index invariant,
since $\mathcal{S}$ switches the upper and lower edges, $\{\mathcal{S},S_z\}=0$, and $[\mathcal{S},C_{ch}]=0$.
As a result, $\gamma^{a,\dagger}_{\lambda=a}$ and $\gamma^{-a,\dagger}_{\lambda=-a}$ are related by $\mathcal{S}$
and are the eigen-operators with the same energy eigenvalue.
Hence, the magnetizations are opposite for the two edges.
As for the case of $\Delta_s<0$, the magnetization at each edge is
reversed with respect to the case of $\Delta_s>0$.
We also note that there is no magnetization in the bulk due to the $PT$-symmetry.

Next we consider the $p\pm is$ superconductors in two dimensions and
show that there appears a single Majorana zero mode localized at the magnetic
kink on the $1D$ edge of this system.
The corresponding BdG Hamiltonian reads
\begin{eqnarray}
H_{2d}&=&\frac{1}{2} \int d^2\vec{r} \psi^\dagger(\vec{r}) \Big\{ \big(-\frac{\hbar^2}{2m}
(\partial_y^2+\partial_z^2)-\mu(z) \big)\tau_3\nn\\
&-& \Delta_s(y) \sigma_2\tau_1 +\frac{1}{k_f}
\big( \Delta_p^y i\partial_y \tau_2 -\Delta_p^z i\partial_z \sigma_1 \tau_1\big)
\Big\}\psi(\vec{r}),
\label{eq:2DHam}~~~
\end{eqnarray}
in which $\vec{r}=(y,z)$,
$\Delta_p^{y,z}$ represent the triplet pairing strengths in the $p_y,\,p_z$ partial wave channels, respectively,
and the $d$-vector is pointing along the momentum direction.
Again an open boundary condition is imposed along the $z$-direction.
With a uniform $\Delta_s(y)$,
the momentum $k_y$ along the $y$-direction is a good quantum number.
The Hamiltonian in Eq. (\ref{eq:2DHam}) reduces back to Eq. (\ref{eq:1DHam})
by setting $k_y=0$, thus there are two gapped modes $\Psi^{a}_{\pm}$
on each edge $a$.

At a small but nonzero $k_y$, the effective $1D$ low energy edge
Hamiltonian can be obtained by the $k\cdot p$ method, as
$H_{2d,edge}^{a}=-a(\Delta_s s_3+\frac{\Delta_p^y}{k_f}k_y  s_1)$,
in which $s_i$'s are the Pauli matrices in the basis of $\Psi^{a}_{\pm}$
for the $a$-edge.
For a spatially slowly varying $\Delta_s(y)$,
the edge Hamiltonian becomes
\begin{equation}
H_{2d,edge}^{a}=-a\left(\Delta_s(y) s_3-i\frac{\Delta_p^y}{k_f}\partial_y  s_1\right).
\label{eq:1DedgeHam}
\end{equation}
Since the direction of the edge magnetization is determined by the sign of $\Delta_s(y)$,
 the position where $\Delta_s(y)$ changes sign forms a magnetic kink separating regions of opposite directions of magnetizations.
Alternatively, $\Delta_s(y)$ can be viewed as the mass of the $1D$ superconducting spinless fermion model,
therefore, a Majorana zero mode emerges at the magnetic kink \cite{Schnyder2008,Qi2008}.
The Majorana zero mode can be solved based on the low energy edge Hamiltonian  in Eq. (\ref{eq:1DedgeHam}).
For a kink with $\text{sgn}(\Delta_s(y))=-\lambda \text{sgn}(y)$ where $\lambda=\pm$, the wavefunction of the zero mode at $a$-edge is
$W^{a}_{\lambda}(y,z)=\frac{1}{\sqrt{2}} \left(e^{ i\lambda \frac{\pi}{4}}\Psi^{a}_{+}(z)+ e^{- i\lambda\frac{\pi}{4}} \Psi^{a}_{-}(z)
\right) w_{\lambda}(y)$,
and the envelope function reads $w_{\lambda}(y)=\frac{1}{N}e^{\lambda \int_0^y dy^{\prime} k_f\frac{\Delta_s(y^{\prime})}{\Delta_p}}$ with $N$ a
normalization factor.
We have also numerically verified the existence of Majorana zero modes using
a lattice model with details included in SM Sect II \cite{suppl}.

\begin{figure}
\includegraphics[width=0.43\textwidth]{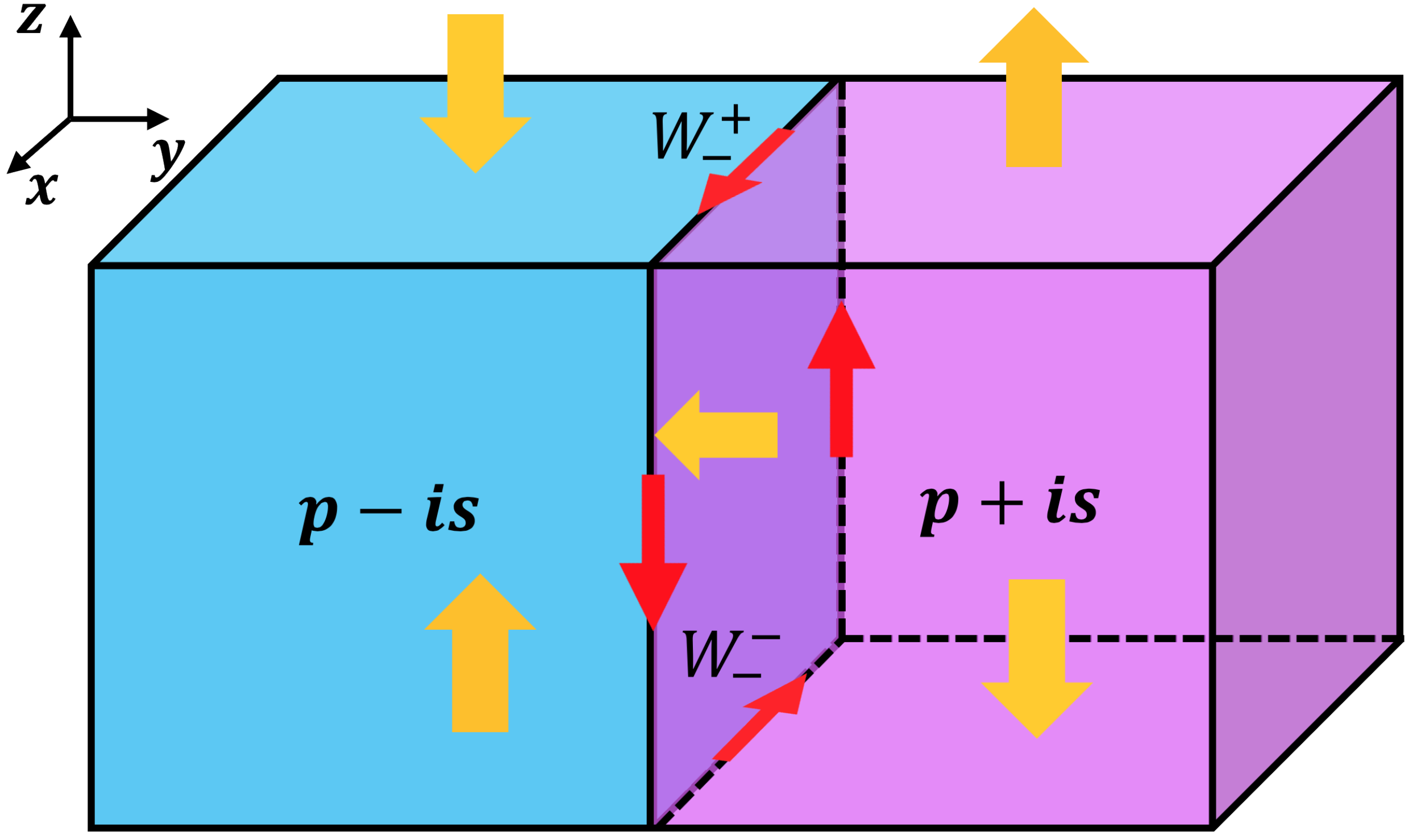}
\caption{The propagating chiral Majorana modes represented by red arrows.
The yellow arrows in the figure represent the directions of magnetizations at several selected surfaces or interface.
}
\label{fig:4zeroModes}
\end{figure}

The symmetry properties of the four zero modes $W^{a}_{\lambda}$'s  are
analyzed as follows.
With the presence of $\Delta_s$, $C_{ch}$ is no longer a symmetry of the zero modes,
nevertheless,  a new chiral operator can be chosen as $C^{\prime}_{ch}=-\sigma_3\tau_1$ which anticommutes with $H_{2d}$
and is a symmetry for $W^{a}_{\lambda}$'s.
When $\Delta_s(y)$ is an odd function of $y$,
the system is invariant under the operation $M_y^{\prime}=GM_y$,
where the reflection operation $M_y$ is defined as $M_y \psi^{\dagger}(y,z)M_y^{-1}=\psi^{\dagger}(-y,z) i\sigma_2\tau_0$.
Furthermore, $C^{\prime}_{ch}$ and $M^{\prime}_y$ commute and form a complete set of good quantum numbers for the four zero modes $W^{a}_{\lambda}$'s:
\begin{eqnarray}
C^{\prime}_{ch} W^{a}_{\lambda}=-a\lambda W^{a}_{\lambda},&
M^{\prime}_y W^{a}_{\lambda}=-\lambda W^{a}_{\lambda}.
\end{eqnarray}
Note that for a fixed $\lambda$, the $C_{ch}^{\prime}$ indices of the pair of states $W^{\pm}_{\lambda}$ are opposite while the $M^{\prime}_y$ eigenvalues are the same,
and for a fixed $a$, both the $C^{\prime}_{ch}$ indices and the $M^{\prime}_y$ eigenvalues of the states $W^{a}_{\pm}$ are opposite.
These are consequences of the symmetries of the system as discussed
in SM Sect. III \cite{suppl}.

Now we extend the above discussions to $3D$ $p\pm is$ superconductors,
with the following BdG Hamiltonian
\begin{eqnarray}
H_{3d}&=& \frac{1}{2}\int d\vec r \psi^\dagger(\vec r) \Big( \big(-\frac{\hbar^2}{2m}
\nabla^2-\mu(z) )\tau_3- \Delta_s(y) \sigma_2\tau_1\nn\\
&& +\frac{1}{k_f}(\Delta^x_p i\partial_x\sigma_3\tau_1+\Delta_p^y i\partial_y \tau_2-\Delta_p^zi\partial_z\sigma_1\tau_1)
\Big)\psi(\vec r),~~~~
\label{eq:3DHam}
\end{eqnarray}
in which $\hat{d}(\vec{k})$ is assumed to be $\hat{k}$.
The open boundary condition is imposed along $z$-direction the same as before.
$H_{3d}$ reduces to the Hamiltonian in Eq. (\ref{eq:1DHam})
when $k_x=k_y=0$, and to Eq. (\ref{eq:2DHam}) when $k_x=0$.

When $\Delta_s(y)=\Delta_s$ is a constant, the surface is uniformly spin polarized.
The surface modes at $k_x=k_y=0$ are $\Psi^{a}_{\lambda}$'s as given in Eq. (\ref{eq:goodquantum1}).
Away from the surface $\Gamma$-point, the low energy surface Hamiltonian for the $a$-boundary can be obtained by the $k\cdot p$ method as
\begin{equation}
H^{a}_{\text{surf}}=a\big(-\Delta_s \xi_3+\frac{1}{k_f} (\Delta_p^x k_x \xi_2-\Delta_p^yk_y\xi_1)\big),
\label{eq:surfHam}
\end{equation}
where $\xi_i$'s are the Pauli matrices in the basis of $\Psi^{a}_{\pm}$.
For the case of rotationally invariant triplet pairing, {\it i.e.} $\Delta_p^j=\Delta_p$ ($j=x,y,z$), the surface magnetization per unit area is evaluated as $a\frac{k_f^2}{8\pi} r (\sqrt{1+r^2} - r)$ where $r=\Delta_s/\Delta_p$,
with detailed calculations included in SM Sect. IV \cite{suppl}.
Due to the surface magnetization, the spontaneous time reversal symmetry breaking pattern (i.e. the sign of $\Delta_s$) can be controlled by an external magnetic field.
By applying  an arbitrarily small field along positive (negative) $z$-direction to the upper boundary of the system above the superconducting transition temperature,
the $p+is$ ($p-is$) state will be favored near the upper boundary when the system is cooled down to be superconducting.
The induced symmetry breaking pattern is the opposite for the lower boundary with the same direction of the external field.

There exists a chiral Majorana fermion propagating along the magnetization domain wall where $\Delta_s(y)$ changes sign on the boundary of the system.
Assuming $\text{sgn}(\Delta_s(y))=-\lambda \text{sgn}(y)$ with $\lambda=\pm 1$,
then for a fixed $\lambda$, there exists a Majorana zero mode $W^{a}_{\lambda}$ with $k_x=0$ on the $a$-boundary.
In this case, a new chiral operator $C^{\prime}_{ch}$ which anti-commutes
with $H_{3d}$ (see SM Sect. III \cite{suppl}) can be defined as $C^{\prime}_{ch}=GM_xTP_h$, satisfying $C^{\prime}_{ch}\psi^{\dagger}(x,y,z)C^{\prime-1}_{ch}=\psi^{\dagger}(-x,y,z) (-\sigma_3\tau_1)$
where $M_x$ is the reflection operation defined as  $M_x \psi^{\dagger}(x,y,z)M_x^{-1}=\psi^{\dagger}(-x,y,z) i\sigma_3\tau_1$.
Since $C^{\prime}_{ch}$ reduces to $-\sigma_3\tau_1$ when $k_x=0$, $C^{\prime}_{ch}$ is a symmetry for the zero modes.
When $k_x$ deviates from 0,
the dispersion can be obtained by applying the $k\cdot p$ method to $\Delta H_{ch}(k_x)=-\frac{\Delta_p^x}{k_f} k_x \sigma_3 \tau_1$,
which is just $\frac{\Delta_p^x}{k_f} C^\prime_{ch} k_x$.
Hence, the propagation direction, {\it i.e.}, the chirality,
is determined by the $C^{\prime}_{ch}$ index and the velocity is
$v=C^{\prime}_{ch}\frac{\Delta^x_p}{\hbar k_f}$.
The above analysis is confirmed by numerical computations on a finite size lattice system discussed in SM Sect. II \cite{suppl}.
A schematic plot of the propagation of the Majorana modes is shown in Fig. \ref{fig:4zeroModes}.

We also show that the spontaneous surface magnetization is a manifestation of a magnetoelectric effect.
In $p\pm is$ superconductors, spatial inhomogeneities, including the
spatial variations of the external potential $V(\vec r)$ and the gap
functions of the $s$ and $p$-wave pairings, can all induce spin polarizations.
In the following, a Ginzburg-Landau free energy analysis is presented
for the mechanism of the magnetization, which holds for temperatures close
to the transition temperature $T_c$ and the slow varying case.
Denote the magnetic field as $\vec{h}(\vec{r})$,
and the singlet and triplet pairing complex gap functions as $\Delta_s(\vec{r})$ and $\Delta_p(\vec{r})$, respectively,
where an isotropic $p$-wave pairing is assumed, {\it i.e.},
$\Delta_p^j=\Delta_p$ ($j=x,y,z$).
The free energy acquires the following terms under $\vec h(\vec r)$ as
\begin{eqnarray}
\Delta F^{(3)}&=&\frac{1}{3}D\epsilon_f \int d^3\vec{r} \,\,\vec{h}\cdot \text{Im} \big[ -(\nabla \Delta_s)\Delta_p^{*}+\Delta_s \nabla \Delta_p^{*} \big],  \nn\\
\Delta F^{(4)}&=&D\int d^3\vec{r}\,\, \vec{h}\cdot \text{Im} \big[ (\nabla V)\Delta_s \Delta^{*}_p -V (\nabla \Delta_s)\Delta^{*}_p\nn\\
&& +V\Delta_s \nabla\Delta_p^{*}\big],
\label{eq:free_energy}
\end{eqnarray}
in which $\epsilon_f$ is the Fermi energy; $k_f$ is the Fermi wavevector;
$D= N_f\frac{1}{k_f} \frac{7\zeta(3)}{(8\pi)^2}\frac{1}{T_c^2}$ where
$N_f$ is the density of states at the Fermi energy,
$T_c$ is the superconducting transition temperature,
and $\zeta$ is the Riemann zeta function.
The derivations of Eq. (\ref{eq:free_energy}) are included in SM
Sect. V\cite{suppl}.
Since the magnetization $\vec M$ is conjugate to $\vec h$,
all $\nabla \Delta_s$, $\nabla\Delta_p$ and $\nabla V$ can induce
$\vec M$.

The first term in $\Delta F^{(4)}$ leads to the magnetoelectric effect.
The magnetization induced by the spatial variation of the external potential
is given by $\vec{M}(\vec{r})=\chi \nabla V(\vec{r})$,
where $\chi=D\text{Im}(\Delta_p\Delta_s^{*})$.
A nonzero $\chi$ requires the coexistence of $\Delta_{s}$ and
$\Delta_{p}$
with a phase difference not equal to $0$ or $\pi$, hence, both the inversion and time-reversal symmetries are broken.
As analyzed before, the $p\pm is$ pairing gap functions are energetically favored when they are nearly degenerate.
In this case, $\chi$ is nonzero and its sign is opposite for the $p\pm is$ cases.
The open boundary condition used previously corresponds to a sudden jump
of the electric potential.
Hence, the spin polarized surface states is a manifestation of
the magnetoelectric effect.

Magnetizations can also be induced by the spatial inhomogeneity of the superconducting gap functions as described in $\Delta F^{(3)}$.
This effect is embodied in the spontaneous magnetization at the interfaces between regions with different pairing symmetries in the bulk as shown in Fig. \ref{fig:4zeroModes} $(b)$.
Furthermore, for the interface between $p+ is$ and $p-is$  regions, we find localized
and spin polarized mid-gap states at the energy of $|\Delta_p|$, while
for the interface between the $ p+is$ and $-p+is$ regions,
the energies  are at $|\Delta_s|$.
The solutions of these mid-gap states are included in SM Sect. VI \cite{suppl}.
We also note that the chiral Majorana fermion can alternatively be viewed as propagating on the edge of such interfaces.

\begin{figure}
        \includegraphics[width=0.47\textwidth]{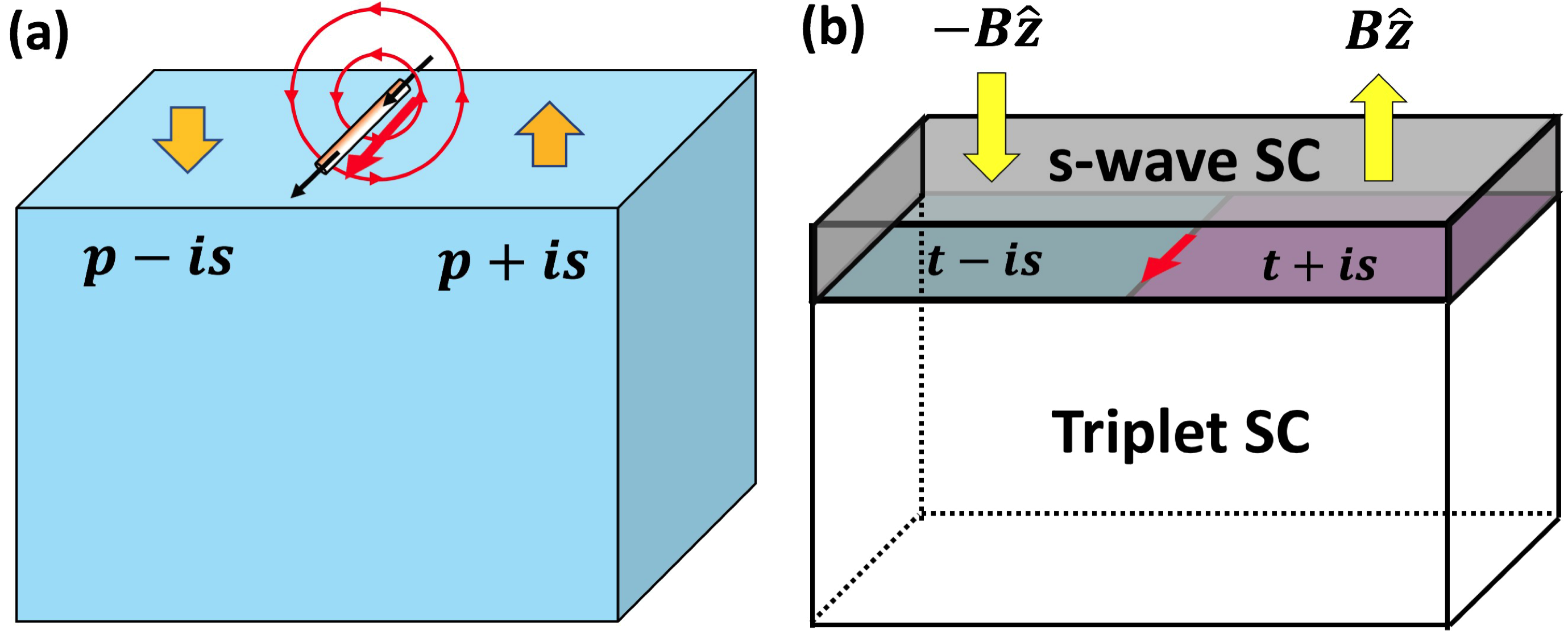}
\caption{The propagating chiral Majorana fermion (represented by the red straight arrows)
produced by
(a) a current-carrying wire placed on top of a $p\pm is$ superconductor,
and (b) a triplet pairing superconductor in proximity with an $s$-wave superconducting thin film.
In (a), the black straight arrows, the red  circled arrows, and the orange fat arrows represent the electrical currents flowing through the wire, the magnetic fields generated by the current, and the surface magnetizations, respectively.
In (b), the yellow arrows represent the directions of the external magnetic fields.
}
\label{fig:proximity}
\end{figure}

The chiral Majorana fermion can be produced and controlled by a current-carrying wire placed on top of the surface of the system as shown in Fig. \ref{fig:proximity} (a).
The directions of the magnetic fields on the surface produced by the wire
are antiparallel on the opposite sides of the wire,
thus the induced symmetry breaking pattern ($p+is$ or $p-is$)
changes across the wire when the system is cooled below $T_c$.
As discussed previously, there exists a chiral Majorana fermion propagating
along the domain wall produced by the electric current.
The domain wall will follow the motion of the wire if the motion is
slow enough to ensure adiabaticity.
Hence, such chiral Majorana fermion is mobile and can be conveniently dragged by translating the wire on the surface.

Besides the intrinsic $p\pm is$ superconductivities,
our strategy for realizing mobile chiral Majorana fermions can also be
carried out using proximity effect.
There is experimental evidence for the Cu$_x$Bi$_2$Se$_3$ and Sn$_{1-x}$In$_x$Te materials
 to host a time reversal invariant triplet superconductivity \cite{Sasaki2011,Kirzhner2012,Sasaki2012,Novak2013}.
In these two materials,  although the proposed triplet pairing symmetry ($A_{1u}$ representation of the $D_{3d}$ point group) \cite{Fu2010}
 is not the same as the $p$-wave pairing discussed in our work,
 $C_{ch}$ is still an anti-symmetry ({\it i.e.}, anti-commuting with the Hamiltonian),
 and $S_z$ is in the little group of $\vec{k}\parallel \hat{z}$ since the $C_3$ symmetry is unbroken.
 Hence, the presence of an $s$-wave pairing will play exactly the same role as discussed in our work
 in splitting the Majorana zero modes and creating a spontaneous magnetization on the surface with the normal direction along $\hat{z}$.
 As shown in Fig. \ref{fig:proximity} (b),
 by coating an $s$-wave superconducting thin film on top of the triplet superconducting bulk,
 the unitary $t\pm is$ pairing symmetry ($t$ for the above mentioned $A_{1u}$ pairing)
 is energetically favored close to the interface between the $s$-wave film and the triplet pairing bulk
 due to the proximity effect.
 The phase difference between the singlet and triplet pairing components ({\it i.e.}, $t+is$ or $t-is$)
 can be conveniently controlled by external magnetic fields,
 and there exists a chiral Majorana fermion propagating along the domain wall
 separating the $t+is$ and the $t-is$ regions on the interface.

In summary, we have proposed that both the localized Majorana zero energy states
and the dispersive chiral ones can be realized via the ``kink on boundary" method.
The boundaries of $p\pm is$ superconductors are spontaneously magnetized,
with opposite directions of magnetizations for the $p+is$ and $p-is$ pairings,
as a manifestation of the magnetoelectric effect.
Along the 1D domain wall between the $p\pm is$ domains on the surface,
there exists a chiral Majorana mode propagating unidirectionally,
which can be controlled by magnetic fields.
Our discussions are relevant to superconducting materials with competing singlet and triplet pairing orders
and proximity-effect-induced superconductivities with a mixed singlet and triplet pairing symmetry.



\let\oldaddcontentsline\addcontentsline
\renewcommand{\addcontentsline}[3]{}


\let\addcontentsline\oldaddcontentsline


\begin{widetext}
\clearpage

\centerline{ {\Large \bf Supplementary Materials} }

\tableofcontents
\section{Surface states on the boundary of $p\pm is$ superconductors}
\label{sec:surface_states}
In this section, we solve for the surface states of a $p+is$ superconductor.

The wavefunction can be written as $(u_{\pm}(z)^T\,v_{\pm}(z)^T)^T e^{i(k_xx+k_yy)}$,
in which both $u_{\pm}(z)$ and $v_{\pm}(z)$ are two-component column vectors,
and plus (minus) sign is for upper (lower) boundary.
Substituting $v_{\pm}(z)$ with $v_{\pm}(z)=\pm i\sigma_1 u_{\pm}(z)$ in the eigen-equation, one obtains
\begin{flalign}
&\pm\big( \frac{\hbar^2}{2m}(-\partial_z^2-k_f^2+k_{\parallel}^2)+ \frac{\Delta_p}{k_f}\partial_z \big) u_{\pm}(z)+\big(-\Delta_s\sigma_3+\frac{\Delta_p}{k_f}(k_x\sigma_2-k_y\sigma_1)\big) u_{\pm}(z)=E_s(\vec{k}_{\parallel})u_{\pm}(z),
\label{eq:surfacestate}
\end{flalign}
in which $\vec{k}_{\parallel}=(k_x,k_y)$, $k_{\parallel}^2=k_x^2+k_y^2$, and $E_s(\vec{k}_{\parallel})$ is the energy of the surface state.
At $k_x=k_y=0$ and $\Delta_s=0$, the equation
\bea
\big( \frac{\hbar^2}{2m}(-\partial_z^2-k_f^2+k_{\parallel}^2)\mp \frac{\Delta_p}{k_f}\partial_z \big) u_{\pm}(z)=0
\eea
has two degenerate solutions as $u_{\pm}(z)=(1\,\,0)^T u^{\pm}_{0}(z)$ and $u_{\pm}(z)=(0\,\,1)^T u^{\pm}_{0}(z)$,
in which $u^{\pm}_{0}(z)$ is given by
\bea
u^{\pm}_{0}(z)=\frac{1}{\sqrt{N(\vec{k}_{\parallel})}}\sin(\sqrt{k_f^2-k_{\parallel}^2} z) e^{\mp \frac{m \Delta_p}{\hbar^2 \sqrt{k_f^2-k_{\parallel}^2}}(z\pm \frac{L}{2})}.
\eea
The energies of these two states will split in the presence of the term $\pm\big(-\Delta_s\sigma_3+\frac{\Delta_p}{k_f}(k_x\sigma_2-k_y\sigma_1)\big) u(z)$ in Eq. (\ref{eq:surfacestate}).
The dispersion is clearly $\pm\sqrt{\Delta_s^2+(\frac{\Delta_p}{k_f})^2k_{\parallel}^2}$.

\section{Numerically solving the Majorana zero modes}

\subsection{The tight binding model for numerical computations}
\label{sec:tight_binding}

In this section, we present the tight binding model on a lattice system which, in the long wavelength limit, reduces to the following Hamiltonian in the main text,
\bea
H_{3d}= \frac{1}{2}\int d\vec r \psi^\dagger(\vec r) \Big( \big(-\frac{\hbar^2}{2m}
\nabla^2-\mu(z) )\tau_3- \Delta_s(y) \sigma_2\tau_1
 +\frac{1}{k_f}(\Delta^x_p i\partial_x\sigma_3\tau_1+\Delta_p^y i\partial_y \tau_2-\Delta_p^zi\partial_z\sigma_1\tau_1)
\Big)\psi(\vec r),~~~~
\label{eq:3DHam}.
\eea
The numerical results displayed in Fig. 1
are based on the model presented in the following.

A two dimensional square lattice is considered with the lattice points at $\vec{R}=a\hat{e}_y+b\hat{e}_z$  ($a,b\in \mathbb{Z}$), where $\hat{e}_y$ (or $\hat{e}_z$) is the unit vector along the $y$ (or $z$)-direction.
For fixed $k_x$, the Hamiltonian of the two dimensional tight binding model reads
\bea
H_t(k_x) &=&  \sum_{\vec{R}} \psi^{\dagger}(k_x,\vec{R}) \big[ -(\mu-k_x^2) \sigma_0\tau_3-\Delta_s(y) \sigma_2\tau_1-k_x\sigma_3\tau_1  \big] \psi(k_x,\vec{R})  \nn\\
&&-t \sum_{\vec{R}}\big[ \big(\psi^{\dagger}(k_x,\vec{R}) \psi(k_x,\vec{R}+\hat{e}_y)+\psi^{\dagger}(k_x,\vec{R}) \psi(k_x,\vec{R}+\hat{e}_y)-2\psi^{\dagger}(k_x,\vec{R}) \psi(k_x,\vec{R})  \big)+h.c.\big]\nn\\
&&+\frac{1}{2i}t \frac{\Delta_p}{\sqrt{\mu}} \sum_{\vec{R}}\big[ \psi^{\dagger}(k_x,\vec{R}) (-\sigma_0\tau_2) \psi(k_x,\vec{R}+\hat{e}_y) +\psi^{\dagger}(k_x,\vec{R}) (\sigma_1\tau_1) \psi(k_x,\vec{R}+\hat{e}_z)+h.c.\big].
\label{eq:tight_binding}
\eea
The Hamiltonian in Eq. (\ref{eq:3DHam}) is translationally invariant along the $x$-direction, hence, $k_x$ is a good quantum number.
It is not hard to see that apart from an overall factor, Eq. (\ref{eq:3DHam}) essentially describes the same system as Eq. (\ref{eq:tight_binding})  in the long wavelength limit by setting $-i\partial_x$ as $k_x$ in Eq. (\ref{eq:3DHam}).


\subsection{Numerical solution}

\begin{figure}
\includegraphics[width=0.3\textwidth]{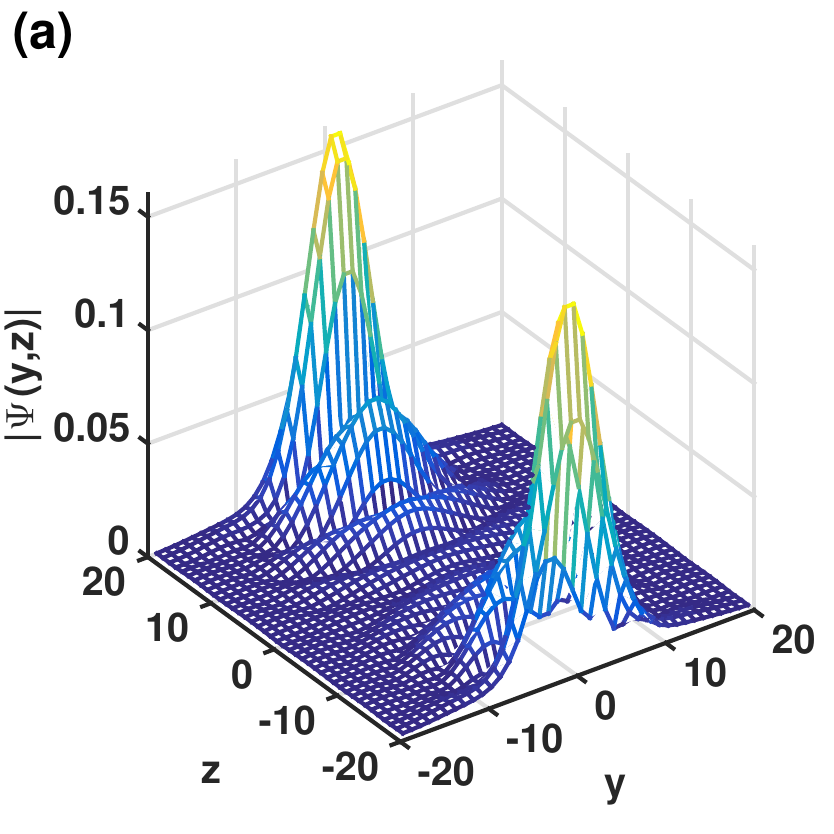}
\includegraphics[width=0.3\textwidth]{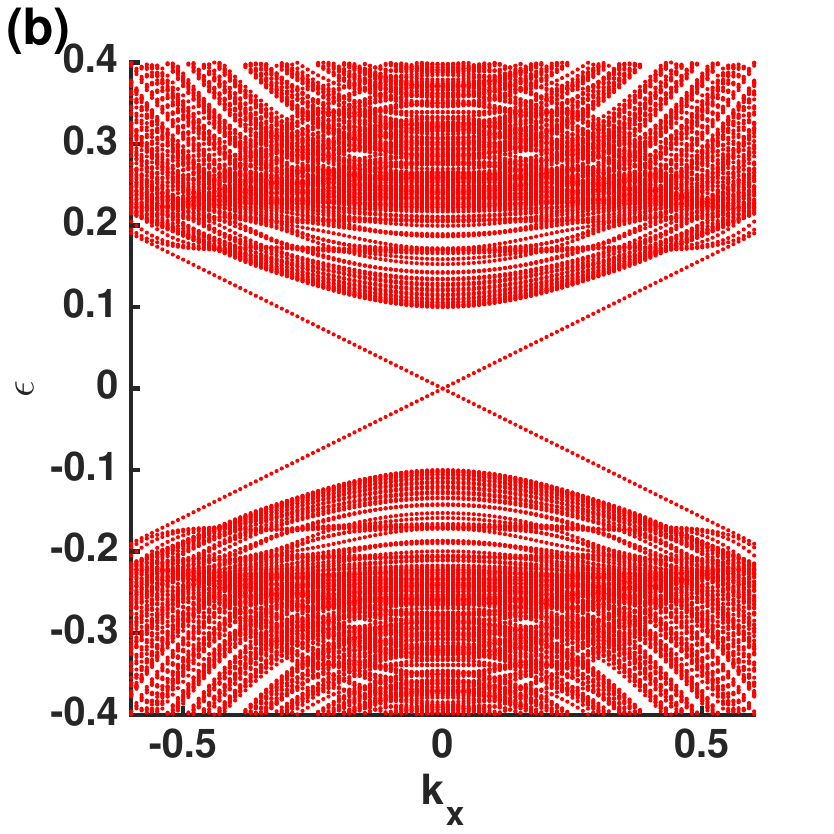}
\caption{$(a)$ The probability density for one of the two zero modes on the 1D edges of a 2D system,
and $(b)$ The spectrum of the Hamiltonian in  Eq. (\ref{eq:3DHam}).
The numerical computation is carried out on a $40\times 40$ lattice system in the $yz$ plane.
In the tight binding model which is used for numerical calculations, the parameters are chosen as $t=1$, $\mu=0.4$, $\Delta_s(y)=-0.1\,\text{sign}(y)$, and $\frac{\Delta_p}{\sqrt{\mu}}=0.2$.
An open boundary condition is taken for both $y$- and $z$-directions.
Due to the finite size effect, there is an energy splitting on the order of $10^{-4}$ between the two zero modes.
}
\label{fig:zero_mode}
\end{figure}

The existence of a Majorana zero mode localized at the magnetic kink
on the 1D boundary of a 2D system is verified by numerical computations as shown in Fig. \ref{fig:zero_mode} $(a)$.
The calculation is performed on a finite size system based on a
tight-binding model with details included in Sec. \ref{sec:tight_binding},
where $\Delta_s(y)=\text{sgn}(y)\Delta_s$.
In Fig. \ref{fig:zero_mode} $(a)$, the wavefunction mixing between the
upper and lower edges is a finite size effect, leading to a small energy splitting of the two zero modes $W^{\pm}_{\lambda}$.
In the thermodynamic limit, the two modes localized at upper and
lower edges are degenerate at zero energy.

\section{Symmetry operations}
\label{sec:sym}
\addtocontents{toc}{\protect\setcounter{tocdepth}{0}}

In this section, we analyze the symmetries of the system and the relations between the symmetry operations.
The coordinate system is chosen according to the convention denoted in Fig. 1 of the main text.

\subsection{$C^{\prime}_{ch}$}

We show that the operation $C^{\prime}_{ch}=GM_xTP_h$ defined in the main text anti-commutes with the Hamiltonian in Eq. (\ref{eq:3DHam}) with $\Delta_s(y)$ an odd function.
The effects of the operations $P_h,T,M_x,G$ on the system are schematically shown in Fig. \ref{fig:Cchprime}.
$P_h$ anti-commutes with the Hamiltonian which is denoted by the overall minus sign in the figure.
The triplet pairing is time reversal invariant, but the singlet component changes sign due to the factor of $i$.
The $s$-wave component is invariant under the reflection operation,
but the $p$-wave component changes sign since $\vec{p}\cdot \vec{\sigma}$ is a pseudo-scalar.
The gauge transformation reverses the sign of the pairing.
As shown in Fig. \ref{fig:Cchprime}, the composed operation $C^{\prime}_{ch}$ anti-commutes with the Hamiltonian.

It can also be explicitly checked that $C_{ch}^{\prime}$ anti-commutes with the Hamiltonian.
The action of $C_{ch}^{\prime}$ is
\bea
C_{ch}^{\prime}&:& (x,y,z)\rightarrow (-x,y,z),(p_x,p_y,p_z)\rightarrow (-p_x,p_y,p_z);\nn\\
&&-\sigma_3\tau_1,
\label{eq:Cchprime}
\eea
in which the first and second row denote how $C_{ch}^{\prime}$ acts in the spatial and internal degrees of freedom, respectively.
$-\sigma_3\tau_1$ anti-commutes with all terms in the Hamiltonian Eq. (\ref{eq:3DHam}) except $-\frac{\Delta_p}{k_f}k_x \sigma_3\tau_1$.
But the spatial part of $C_{ch}^{\prime}$ reverses the sign of $k_x$.
Hence, $\{C_{ch}^{\prime},H\}=0$.

\begin{figure}
\includegraphics[width=0.65\textwidth]{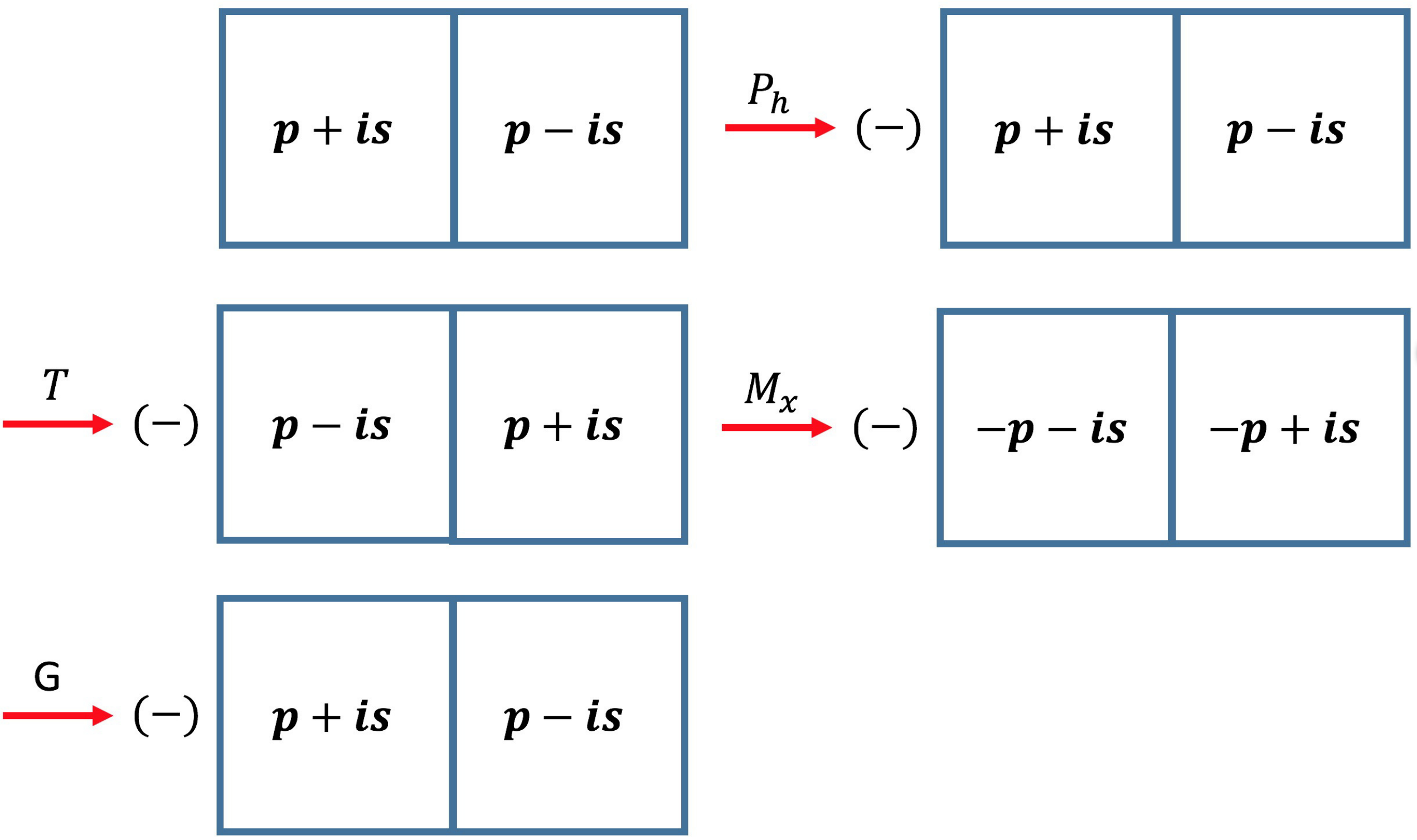}
\caption{The symmetry operation $C^{\prime}_{ch}=GM_xTP_h$.
}
\label{fig:Cchprime}
\end{figure}

\begin{figure}
\includegraphics[width=0.61\textwidth]{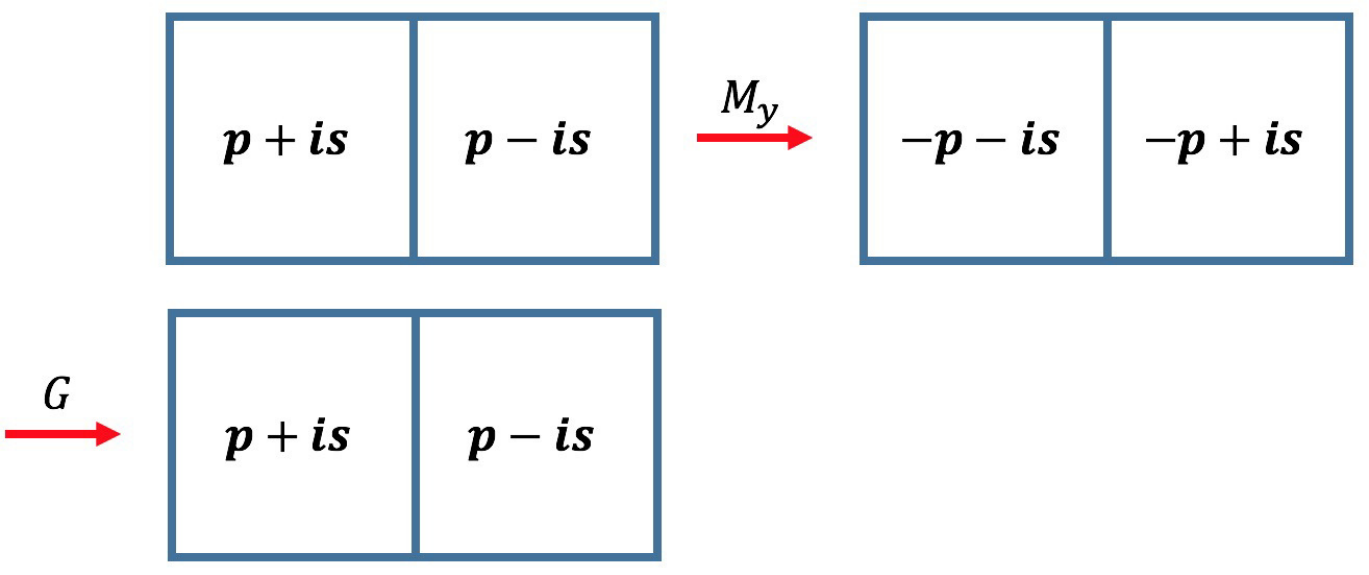}
\caption{The symmetry operation $M^{\prime}_y=GM_y$.
}
\label{fig:Myprime}
\end{figure}

\begin{figure}
\includegraphics[width=0.6\textwidth]{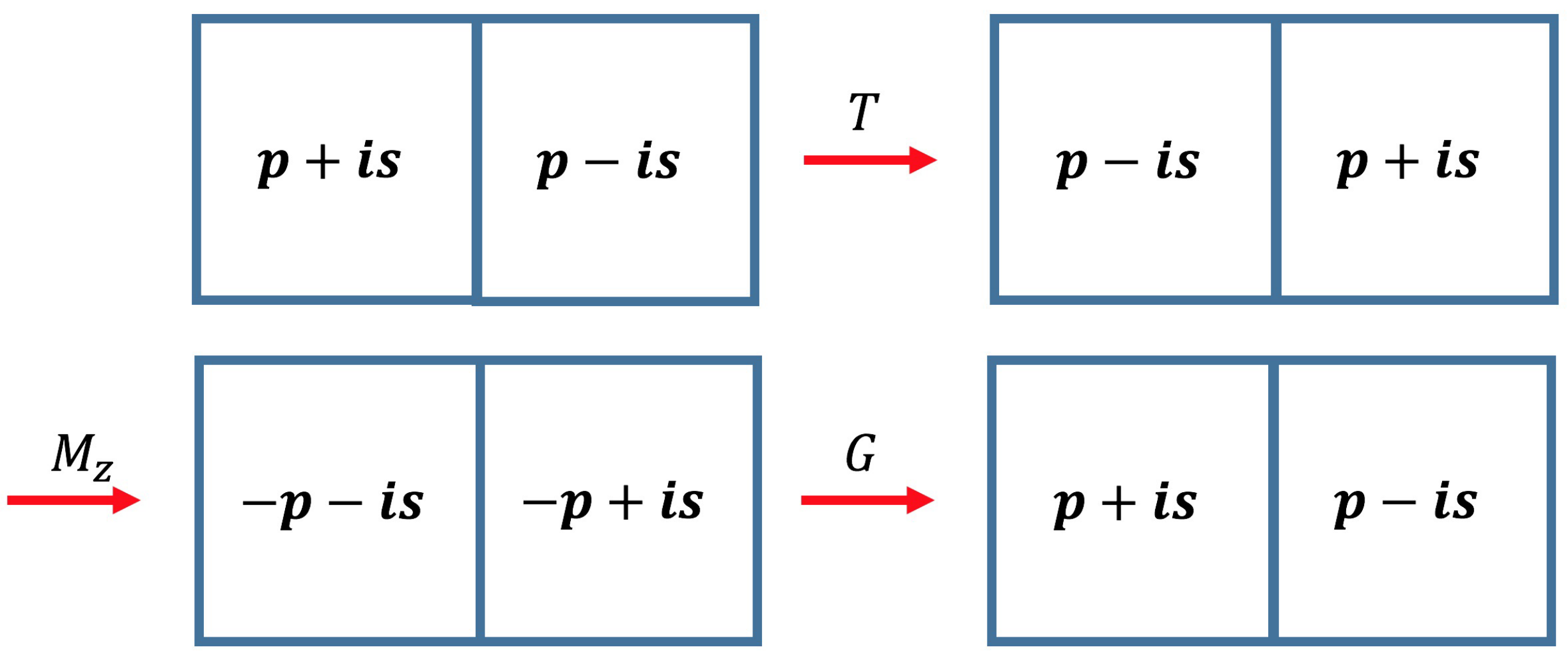}
\caption{The symmetry operation $\bar{M}_z=GM_zT$.
}
\label{fig:Mzbar}
\end{figure}

\subsection{$M^{\prime}_y$}

We now show that the operation $M^{\prime}_y=GM_y$ defined in the main text commutes with the Hamiltonian in Eq. (\ref{eq:3DHam}).
The effects of the operations $M_y,G$ on the system are schematically shown in Fig. \ref{fig:Myprime}.
$M_y$ interchanges the left and right bulks and keeps the $s$-wave component invariant.
$M_y$ also reverses the sign of the $p$-wave component since $\vec{p}\cdot \vec{\sigma}$ is a pseudo-scalar.
The gauge transformation changes the sign of the pairing.
 Thus the operation $M^{\prime}_y$ brings the system back.

It can also be explicitly checked that $M^{\prime}_y$ commutes with the Hamiltonian.
The action of $M_y^{\prime}$ is
\bea
M_{y}^{\prime}&:& (x,y,z)\rightarrow (x,-y,z),(p_x,p_y,p_z)\rightarrow (p_x,-p_y,p_z);\nn\\
&&-\sigma_2\tau_3.
\label{eq:Myprime}
\eea
$-\sigma_2\tau_3$ commutes with all terms in the Hamiltonian in Eq. (\ref{eq:3DHam}) except for $-\frac{\Delta_p}{k_f} k_y\sigma_0\tau_2$ and $-\Delta_s(y) \sigma_2\tau_1$.
But the spatial part of $M_y^{\prime}$ reverses the sign of both these two terms.
Thus $[M_y^{\prime},H]=0$.

\subsection{$\bar{M}_z$}

In this part, we show that the operation $\bar{M}_z=GM_zT$
commutes with $H$ and $M^{\prime}_y$, and anti-commutes with $C^{\prime}_{ch}$.
The effects of the operations $T,M_z,G$ on the system are schematically shown in Fig. \ref{fig:Mzbar},
and the combination of them keeps the system invariant.
The commutation relations can be checked explicitly.
The action of $\bar{M}_z$ is
\bea
\bar{M}_z&:& (x,y,z)\rightarrow (x,y,-z),(p_x,p_y,p_z)\rightarrow (-p_x,-p_y,p_z);\nn\\
&&\sigma_1\tau_0 K.
\label{eq:Mzbar}
\eea
$\sigma_1\tau_0 K$ commutes with all terms in the Hamiltonian except for $-\frac{\Delta_p}{k_f} p_x \sigma_3\tau_1$ and $-\frac{\Delta_p}{k_f} p_y\sigma_0 \tau_2$,
but the spatial part of the action of $\bar{M}_z$ reverses the sign of $p_x,p_y$,
thus $[\bar{M}_z,H]=0$.
Using the explicit expressions in Eqs. (\ref{eq:Cchprime},\ref{eq:Myprime},\ref{eq:Mzbar}), the relations $\{\bar{M}_z,C^{\prime}_{ch}\}=0$ and $[\bar{M}_z,M^{\prime}_y]=0$ can be easily checked.

\subsection{$T$}

In this part, we show that $T$ switches the bulks of $p+is$ and $p-is$ pairings and satisfies the anti-commutation relations $\{T,C^{\prime}_{ch}\}=0$ and $\{T,M^{\prime}_y\}=0$.
The action of the time reversal operation is plotted in the first arrow of Fig. \ref{fig:Mzbar}.
Clearly $T$ interchanges the left and right bulks.
The explicit form of the action of $T$ is
\bea
T&:&(x,y,z)\rightarrow(x,y,z),(p_x,p_y,p_z)\rightarrow (-p_x,-p_y,-p_z);\nn\\
&&i\sigma_2 K.
\label{eq:T}
\eea
Using the expressions in Eqs. (\ref{eq:Cchprime},\ref{eq:Myprime},\ref{eq:T}), it can be verified that
$\{T,C^{\prime}_{ch}\}=0$, $\{T,M^{\prime}_y\}=0$.

\subsection{Symmetries of the Majorana zero modes}

\begin{figure}
\includegraphics[width=0.7\textwidth]{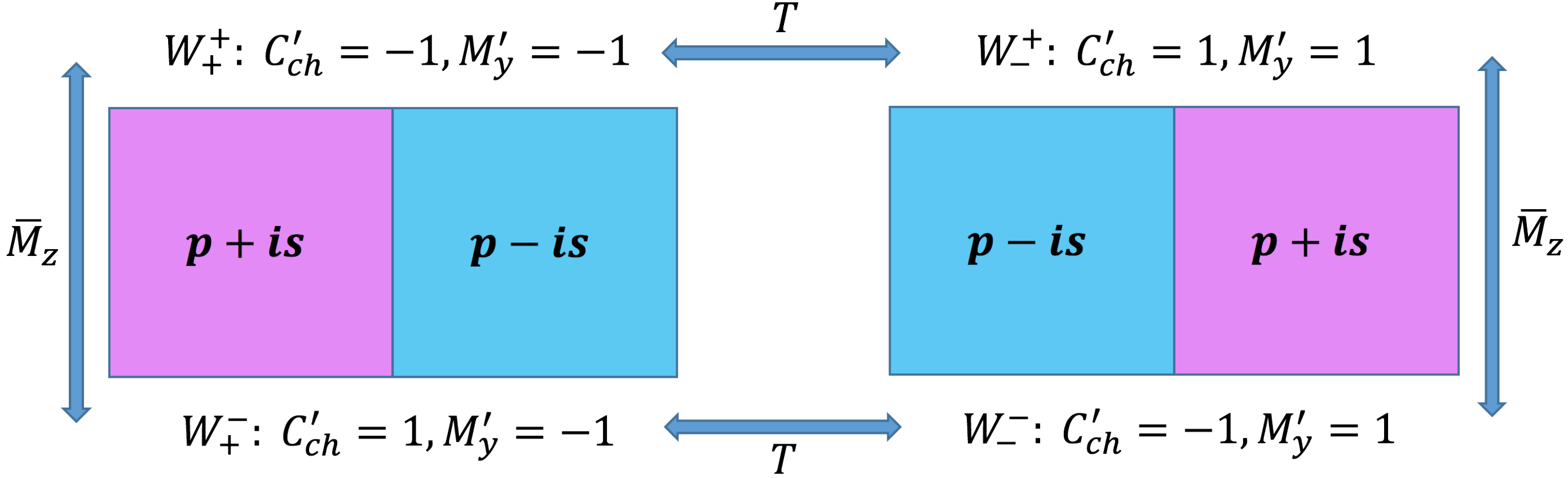}
\caption{Symmetry operations relating the Majorana zero modes at the ``kinks on boundaries".
}
\label{fig:4zeroModes}
\end{figure}

For a fixed $\lambda$, the $C_{ch}^{\prime}$ indices of the pair of states $W^{\pm}_{\lambda}$ are opposite, while the $M^{\prime}_y$ eigenvalues are the same,
which is the consequence of the symmetry operation $\bar{M}_z=GM_zT$,
where $M_z$ is the reflection operation defined as $M_z\psi^{\dagger}(y,z)M_z^{-1}=\psi^{\dagger}(y,-z) i\sigma_3\tau_3$.
$W^{a}_{\lambda}$ and $W^{-a}_{\lambda}$ are related by $\bar{M}_z$ with opposite $C^{\prime}_{ch}$ indices and same $M_y^{\prime}$ eigenvalues,
since $\bar{M}_z$ switches the upper and lower edges, and $[\bar{M}_z,M^{\prime}_y]=0$ and $\{\bar{M}_z,C^{\prime}_{ch}\}=0$.
For a fixed $a$, both the $C^{\prime}_{ch}$ indices and the $M^{\prime}_y$ eigenvalues of the states $W^{a}_{\pm}$ are opposite.
This is due to the time reversal operation,
which switches the $p+is$ and $p-is$ bulks
and anti-commutes with both $C^{\prime}_{ch}$ and $M^{\prime}_y$.
The relations between the $C_{ch}^{\prime}$ indices and the $M_y^{\prime}$ eigenvalues of the four Majorana modes is schematically shown in Fig. \ref{fig:4zeroModes} $(a)$.

\section{Surface magnetization of $p\pm is$ superconductors}

\subsection{Calculation of surface magnetization in $p\pm is$ superconductors}
\label{sec:surface_magnetization}

In this section, we compute the spontaneous magnetization on the boundary of the $p+is$ superconductor.
The upper boundary is taken as an example for calculations.

The quasiparticle  creation operator can be expressed as
\bea
\gamma^{\dagger}_{a,\vec{k}_{\parallel}} =\int dz \psi^{\dagger}(\vec{k}_{\parallel},z) \Psi^{+}_{a}(\vec{k}_{\parallel},z),
\eea
in which $a=\pm$.
The second quantized form of the surface Hamiltonian is
\bea
H_{\text{surf}} = \frac{1}{2} \sum_{\vec{k}_{\parallel}} (\gamma^{\dagger}_{+,\vec{k}_{\parallel}} \,\,\,\gamma^{\dagger}_{-,\vec{k}_{\parallel}}) H_{\text{surf}}(\vec{k}_{\parallel})
\left(\begin{array}{c}
\gamma_{+,\vec{k}_{\parallel}}\\
\gamma_{-,\vec{k}_{\parallel}}
\end{array}\right),
\label{eq:surf2ndquantized}
\eea
in which the matrix kernel $H_{\text{surf}}(\vec{k}_{\parallel})$ is given by $H_{\text{surf}}(k_x,k_y)=-\vec{h}(\vec{k}_{\parallel})\cdot \vec{\xi}$,
where $\vec{h}(\vec{k}_{\parallel})=(\frac{\Delta_p}{k_f} k_y,-\frac{\Delta_p}{k_f} k_x,\Delta_s)^T$,
$\xi_i$'s are the Pauli matrices,
and $\vec{\xi}=(\xi_1,\xi_2,\xi_3)^T$.
Due to the relation
\bea
\gamma_{+,\vec{k}_{\parallel}}^{\dagger}=\gamma_{-,-\vec{k}_{\parallel}},
\eea
the Hamiltonian in Eq. (\ref{eq:surf2ndquantized}) can be written as
\bea
H_{\text{surf}} = \frac{1}{2} \sum_{\vec{k}_{\parallel}} (\gamma_{-,-\vec{k}_{\parallel}} \,\,\,\gamma^{\dagger}_{-,\vec{k}_{\parallel}}) H_{\text{surf}}(\vec{k}_{\parallel})
\left(\begin{array}{c}
\gamma^{\dagger}_{-,-\vec{k}_{\parallel}}\\
\gamma_{-,\vec{k}_{\parallel}}
\end{array}\right).
\eea
In the following, we will drop the subscript ``$-$" and simply write $\gamma_{-,\vec{k}_{\parallel}}$ as $\gamma_{\vec{k}_{\parallel}}$.

Define $a_{\vec{k}_{\parallel}}$ through the relation
\bea
\left(\begin{array}{c}
\gamma_{-\vec{k}_{\parallel}}^{\dagger}\\
\gamma_{\vec{k}_{\parallel}}
\end{array}\right)
=U(\hat{h})
\left(\begin{array}{c}
a_{-\vec{k}_{\parallel}}^{\dagger}\\
a_{\vec{k}_{\parallel}}
\end{array}\right)
\eea
in which $U(\hat{h})=e^{-i\frac{1}{2} \xi_z \phi_h} e^{-i\frac{1}{2} \xi_y \theta_h}$,
where $\theta_h$ and $\phi_h$ are the polar and azimuthal angles of the vector $\vec{h}(\vec{k}_{\parallel})$.
The surface Hamiltonian becomes diagonal in terms of $a_{\vec{k}_{\parallel}}$ as
\bea
H_{\text{surf}}=-\frac{1}{2} \sqrt{\Delta_s^2+\frac{\Delta_p^2}{k_f^2} k_{\parallel}^2} \sum_{\vec{k}_{\parallel}} (a_{-\vec{k}_{\parallel}}\,\,\, a_{\vec{k}_{\parallel}}^{\dagger})  \xi_3
\left(\begin{array}{c}
a_{-\vec{k}_{\parallel}}^{\dagger}\\
a_{\vec{k}_{\parallel}}
\end{array}\right).\nn\\
\label{eq:2ndquantized_Hsurf}
\eea

Projecting the spin operator in $z$-direction to the subspace of the
surface states, one obtains
\bea
S^{z}=\frac{1}{4} \sum_{\vec{k}_{\parallel}} (\gamma_{-\vec{k}_{\parallel}} \,\,\,\gamma^{\dagger}_{\vec{k}_{\parallel}}) \xi_3
\left(\begin{array}{c}
\gamma_{-\vec{k}_{\parallel}}^{\dagger}\\
\gamma_{\vec{k}_{\parallel}}
\end{array}\right).
\eea
In terms of $a_{\vec{k}_{\parallel}}$, the $S_z$ operator
becomes
\bea
S^z=\frac{1}{4} \sum_{\vec{k}_{\parallel}} (a_{-\vec{k}_{\parallel}}\,\,\, a_{\vec{k}_{\parallel}}^{\dagger}) \Lambda(\vec{k}_{\parallel})\cdot \vec{\xi} \left(\begin{array}{c}
a_{-\vec{k}_{\parallel}}^{\dagger}\\
a_{\vec{k}_{\parallel}}
\end{array}\right),
\eea
in which $\Lambda(\vec{k}_{\parallel})=\frac{1}{\sqrt{\Delta_s^2+\frac{\Delta_p^2}{k_f^2}k_{\parallel}^2}} (-\frac{\Delta_p}{k_f}k_{\parallel},0,\Delta_s )$.

The ground state $\ket{G}$ of the Hamiltonian Eq. (\ref{eq:2ndquantized_Hsurf}) is annihilated by $a_{\vec{k}_{\parallel}}$.
Hence
\bea
M_z=\frac{1}{A}\bra{G}S^z\ket{G}=\frac{1}{4} \int^{k_f} \frac{d^2\vec{k}_{\parallel}}{(2\pi)^2} \frac{\Delta_s}{\sqrt{\Delta_s^2+\frac{\Delta_p^2}{k_f^2}k_{\parallel}^2} },
\label{eq:Mz}
\eea
in which $A$ is the area of the system.
The integral is evaluated to be
\bea
M_z=\frac{k_f^2}{8\pi} \frac{\Delta_s}{\Delta_p} (\sqrt{(1+(\frac{\Delta_s}{\Delta_p})^2 }-\frac{\Delta_s}{\Delta_p} ).
\label{eq:result_Mz}
\eea

\section{Ginzburg-Landau free energy analysis of the magnetoelectric effect}

\subsection{Evaluation of $\Delta F^{(3)}$}
\label{sec:free_energy3}

\begin{figure}
\includegraphics[width=0.4\textwidth]{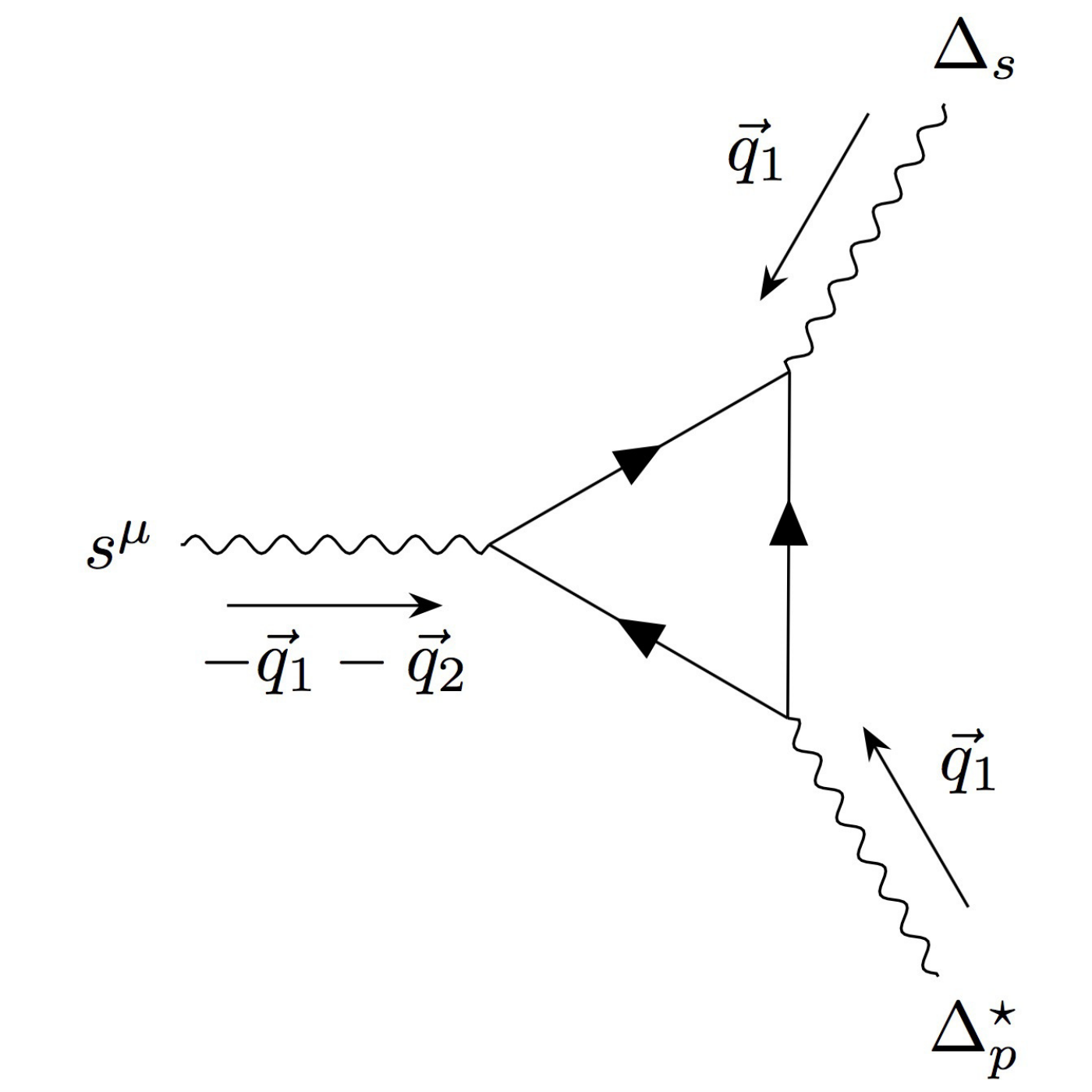}
\caption{The diagram contributing to $\Delta F^{(3)}$ with the combination $\Delta_s\Delta_p^{*}$.
}
\label{fig:diagram3}
\end{figure}

There are two diagrams contributing to $\Delta F^{(3)}$.
The one with the combination $\Delta_s\Delta_p^{*}$ is shown in Fig. \ref{fig:diagram3}.
The other one with the combination $\Delta^{*}_s\Delta_p$ can be obtained by taking the complex conjugate.
Denote $\mathcal{C}_{\mu}$ to be the value of the diagram in Fig. \ref{fig:diagram3}.
In this section, we evaluate $\mathcal{C}$ up to linear order of $\vec{q}_{1},\vec{q}_{2},\vec{q}_{3}$ in the static limit.
We work in the imaginary time formalism, and $\hbar$ is set to be $1$.
The Fermi energy and Fermi wave vector are denoted as $\epsilon_f$ and $k_f$.

The expression for $\mathcal{C}_{\mu}$ is
\bea
\mathcal{C}_{\mu}&=& \int \frac{d^3\vec{k}}{(2\pi)^3} \frac{1}{\beta} \sum_{i\omega_n}
\text{Tr} \big[ \frac{1}{-i\omega_n-\xi_{\vec{k}}} i\sigma_2 \frac{1}{i\omega_n-\xi_{-\vec{k}-\vec{q}_2}} \frac{1}{2} \sigma_{\mu} \frac{1}{i\omega_n-\xi_{-\vec{k}+\vec{q}_1}}\frac{1}{k_f} (\vec{k}-\frac{\vec{q}_1}{2})\cdot \vec{\sigma} i\sigma_2
\big] \nn\\
&=&\frac{1}{k_f} \int \frac{d^3\vec{k}}{(2\pi)^3} \frac{1}{\beta} \sum_{i\omega_n} (k_{\mu}-\frac{q_{1\mu}}{2}) \frac{(-i\omega_n+\xi_{\vec{k}}) (i\omega_n+\xi_{\vec{k}+\vec{q}_2}) (i\omega_n+\xi_{\vec{k}-\vec{q}_1}) }{(\omega_n^2+\xi_{\vec{k}}^2)(\omega_n^2+\xi_{\vec{k}+\vec{q}_2}^2)  (\omega_n^2+\xi_{\vec{k}-\vec{q}_1}^2) }.
\label{eq:C}
\eea

Denote $\mathcal{C}_{\mu}^{(0)}$ as the $\vec{q}_1=\vec{q}_2=\vec{q}_3=0$ term, $\mathcal{C}_{\mu}^{(1)}$ as the linear in $\vec{q}_1$ term, and $\mathcal{C}_{\mu}^{(2)}$ as the linear in $\vec{q}_2$ term of Eq. (\ref{eq:C}).
We obtain
\bea
\mathcal{C}_{\mu}^{(0)}&=&\frac{1}{k_f} \int \frac{d^3\vec{k}}{(2\pi)^3}
\frac{1}{\beta} \sum_{i\omega_n} k_{\mu}
\frac{i\omega_n+\xi_{\vec{k}}}{(\omega_n^2+\xi_{\vec{k}})^2}\nn\\
&=&0,
\eea
\bea
\mathcal{C}_{\mu}^{(1)}&=&\frac{1}{k_f} \int \frac{d^3\vec{k}}{(2\pi)^3}
\frac{1}{\beta} \sum_{i\omega_n} k_{\mu} v_{\vec{k}} \hat{k}\cdot \vec{q}_1
\frac{(i\omega_n+\xi_{\vec{k}})^2}{(\omega_n^2+\xi_{\vec{k}}^2)^3} \nn\\
&=&\frac{1}{3} v_f q_{1\mu} N_f \int d\epsilon \frac{1}{\beta} \sum_{i\omega_n} \frac{\epsilon^2-\omega_n^2}{(\omega_n^2+\epsilon^2)^3},
\eea
\bea
\mathcal{C}_{\mu}^{(2)}&=&-\frac{1}{k_f} \int \frac{d^3\vec{k}}{(2\pi)^3} \frac{1}{\beta} \sum_{i\omega_n} k_{\mu} v_{\vec{k}} \hat{k}\cdot \vec{q}_2 \frac{(i\omega_n+\xi_{\vec{k}})^2}{(\omega_n^2+\xi_{\vec{k}}^2)^3}\nn\\
&=&-\frac{1}{3} v_f q_{2\mu} N_f \int d\epsilon \frac{1}{\beta} \sum_{i\omega_n} \frac{\epsilon^2-\omega_n^2}{(\omega_n^2+\epsilon^2)^3}.
\eea
Denote $C^{(i)}$ ($i=1,2$) to be the coefficient of $q_{i\mu}$ in $\mathcal{C}_{\mu}^{(i)}$.
Using
\bea
\int d\epsilon \frac{1}{\beta} \sum_{i\omega_n} \frac{\epsilon^2}{(\omega_n^2+\epsilon^2)^3}&=&\frac{1}{3}\int d\epsilon \frac{1}{\beta} \sum_{i\omega_n} \frac{\omega_n^2}{(\omega_n^2+\epsilon^2)^3}=\frac{1}{4} \frac{7\zeta(3)}{(8\pi)^2}\frac{1}{T^2},
\eea
we obtain
\bea
-C^{(1)}=C^{(2)}=\frac{1}{6} N_f v_f \frac{7\zeta(3)}{(8\pi)^2} \frac{1}{T^2}.
\eea
When the system is close to the superconducting transition point, we can set $T=T_c$.
The diagram in Fig. \ref{fig:diagram3} represents six terms in the Trlog-expansion of the free energy.
Since Fig. \ref{fig:diagram3} is a third order term, there is an additional $\frac{1}{3}$ factor.
Combing these together, we arrive at the expression of $\Delta F^{(3)}$.

\subsection{Evaluation of $\Delta F^{(4)}$}
\label{sec:free_energy4}

\begin{figure*}
\includegraphics[width=1.0\textwidth]{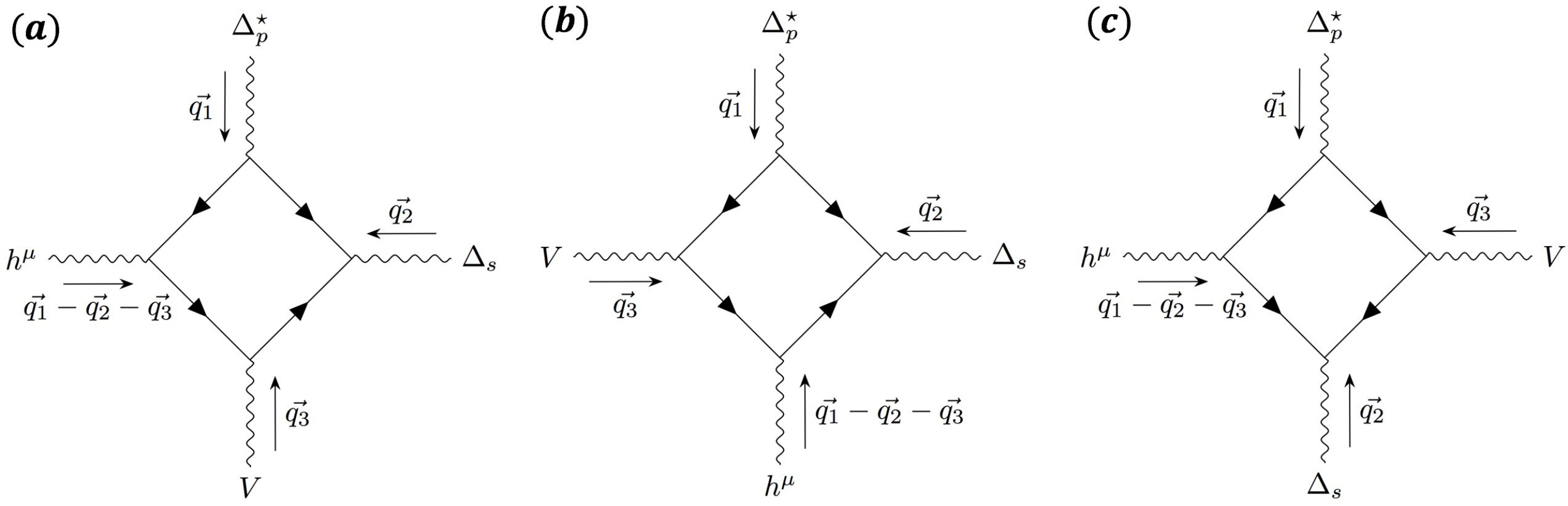}
\caption{The three diagrams contributing to $\Delta F^{(4)}$ with the combination $\Delta_s\Delta_p^{*}$.
}
\label{fig:diagram}
\end{figure*}

There are six diagrams contributing to $\Delta F^{(4)}$.
Three of them with the combination $\Delta_s\Delta_p^{*}$ are shown in Fig. \ref{fig:diagram}.
The other three with the combination $\Delta^{*}_s\Delta_p$ can be obtained by taking the complex conjugate.
Let $\mathcal{D}^{a,b,c}_{\mu}$ be the values of the diagrams $(a,b,c)$ in Fig. \ref{fig:diagram}, respectively.
In this section, we evaluate $\mathcal{D}^{a,b,c}_{\mu}$ up to linear order of $\vec{q}_{1},\vec{q}_{2},\vec{q}_{3}$ in the static limit.

First consider the diagram Fig. \ref{fig:diagram} $(a)$.
The expression for $\mathcal{D}^a_{\mu}$ is
\bea
\mathcal{D}^a_{\mu}&=&\int \frac{d^3\vec{k}}{(2\pi)^3} \frac{1}{\beta} \sum_{i\omega_n}
\text{Tr}\big[
\frac{1}{-i\omega_n-\xi_{\vec{k}}}i\sigma_2\frac{1}{i\omega_n-\xi_{-\vec{k}-\vec{q}_2}}\frac{1}{i\omega_n-\xi_{-\vec{k}-\vec{q}_2-\vec{q}_3}}\frac{1}{2} \sigma_{\mu} \frac{1}{i\omega_n-\xi_{-\vec{k}+\vec{q}_1}}\frac{1}{k_f} (\vec{k}-\frac{\vec{q}_1}{2} )\cdot \vec{\sigma}i\sigma_2
\big]\nn\\
&=& -\int \frac{d^3\vec{k}}{(2\pi)^3} \frac{1}{\beta} \sum_{i\omega_n}
\frac{1}{k_f}(k_{\mu}-\frac{q_{1\mu}}{2}) \frac{(-i\omega_n+\xi_{\vec{k}}) (i\omega_n+\xi_{\vec{k}+\vec{q}_2}) (i\omega_n+\xi_{\vec{k}+\vec{q}_2+\vec{q}_3 }) (i\omega_n+\xi_{\vec{k}-\vec{q}_1}) }{(\omega_n^2+\xi_{\vec{k}}^2) (\omega_n^2+\xi_{\vec{k}+\vec{q}_2}^2) (\omega_n^2+\xi_{\vec{k}+\vec{q}_2+\vec{q}_3}^2) (\omega_n^2+\xi_{\vec{k}-\vec{q}_1}^2) },
\label{eq:D_a}
\eea
in which $\xi_{\vec{k}}=\frac{k^2}{2m}-\epsilon_f$, $\frac{1}{2}\sigma_{\mu}$ is the vertex for the spin operator,
$i\sigma_2$ is the vertex for the singlet pairing,
and $\frac{1}{k_f}(\vec{k}-\frac{\vec{q}_1}{2})\cdot \vec{\sigma}i\sigma_2$ is the vertex for the triplet pairing.
Denote $\mathcal{D}^{a(0,1,2,3)}_{\mu}$ to be the $\vec{q}_1=\vec{q}_2=\vec{q}_3=0$ term, the linear in $\vec{q}_1$ term, the linear in $\vec{q}_2$ term, and the linear in $\vec{q}_3$ term of Eq. (\ref{eq:D_a}), respectively.

For $\mathcal{D}^{a(0)}_{\mu}$, we obtain
\bea
\mathcal{D}^{a(0)}_{\mu}=-\frac{1}{k_f}\int \frac{d^3\vec{k}}{(2\pi)^3} \frac{1}{\beta} \sum_{i\omega_n}
k_{\mu} \frac{(i\omega_n+\xi_{\vec{k}})^2}{(\omega_n^2+\xi_{\vec{k}}^2)^3}=0,
\eea
due to the rotational invariance of the dispersion $\xi_{\vec{k}}$.

For $\mathcal{D}^{a(1)}_{\mu}$, up to linear order in $\vec{q}_1$, we obtain
\bea
\mathcal{D}^{a(1)}_{\mu}=\mathcal{D}^{a(1),1}_{\mu}+\mathcal{D}^{a(1),2}_{\mu},
\eea
in which
\bea
\mathcal{D}^{a(1),1}_{\mu}&=& -\frac{1}{k_f} \int \frac{d^3\vec{k}}{(2\pi)^3} \frac{1}{\beta} \sum_{i\omega_n}
k_{\mu} v_k \hat{k}\cdot \vec{q}_1 \frac{(i\omega_n+\xi_{\vec{k}})^3}{(\omega_n^2+\xi_{\vec{k}}^2)^4}\nn\\
&=& - \frac{1}{3} \frac{q_{1\mu}}{mk_f} \int \frac{d^3\vec{k}}{(2\pi)^3} \frac{1}{\beta} \sum_{i\omega_n}k^2
\frac{(i\omega_n+\xi_{\vec{k}})^3}{(\omega_n^2+\xi_{\vec{k}}^2)^4},
\eea
\bea
\mathcal{D}^{a(1),2}_{\mu}&=&\frac{q_{1\mu}}{k_f} \int \frac{d^3\vec{k}}{(2\pi)^3} \frac{1}{\beta} \sum_{i\omega_n} \frac{(i\omega_n+\xi_{\vec{k}})^2}{(\omega_n^2+\xi_{\vec{k}}^2)^3}.
\eea
To the lowest order in the expansion of $\frac{1}{\beta \epsilon_f}$,
we can use the approximations $\int \frac{d^3\vec{k}}{(2\pi)^3}=N_f\int d\epsilon (1+\frac{\epsilon}{2\epsilon_f})$ and $k=k_f+\frac{\xi_{\vec{k}}}{v_f}$, where $N_f$ is the density of states at Fermi energy.
Then one obtains
\bea
\mathcal{D}^{a(1),1}_{\mu}&=&-N_f\frac{q_{1,\mu}}{k_f} \int d\epsilon \frac{1}{\beta} \sum_{i\omega_n} \frac{\epsilon^2(\epsilon^2-3\omega_n^2)}{(\omega_n^2+\epsilon^2)^4},
\eea
\bea
\mathcal{D}^{a(1),2}_{\mu}&=& \frac{1}{2}N_f \frac{q_{1\mu}}{k_f} \int d\epsilon \frac{1}{\beta} \sum_{i\omega_n} \frac{\epsilon^2-\omega_n^2}{(\omega_n^2+\epsilon^2)^2}.
\eea

For $\mathcal{D}_{\mu}^{a(2)}$ and $\mathcal{D}_{\mu}^{a(3)}$, the procedure of the evaluation is similar.
Up to linear order in $\vec{q}_2$ and $\vec{q}_3$, we obtain,
\bea
\mathcal{D}_{\mu}^{a(2)}&=& 2N_f \frac{q_{2\mu}}{k_f} \int d\epsilon \frac{1}{\beta} \sum_{i\omega_n} \frac{\epsilon^2(\epsilon^2-3\omega_n^2)}{(\omega_n^2+\epsilon^2)^4},
\eea
\bea
\mathcal{D}_{\mu}^{a(3)}&=& N_f \frac{q_{3\mu}}{k_f} \int d\epsilon \frac{1}{\beta} \sum_{i\omega_n}\frac{\epsilon^2(\epsilon^2-3\omega_n^2)}{(\omega_n^2+\epsilon^2)^4}.
\eea

Next consider the diagram Fig. \ref{fig:diagram} $(b)$.
The expression for $\mathcal{D}^b_{\mu}$ is
\bea
\mathcal{D}^b_{\mu}&=&\int \frac{d^3\vec{k}}{(2\pi)^3} \frac{1}{\beta} \sum_{i\omega_n}
\text{Tr}\big[
\frac{1}{-i\omega_n-\xi_{\vec{k}}}i\sigma_2\frac{1}{i\omega_n-\xi_{-\vec{k}-\vec{q}_2}}  \frac{1}{2}\sigma_{\mu} \frac{1}{i\omega_n-\xi_{-\vec{k}+\vec{q}_1+\vec{q}_3}}
\frac{1}{i\omega_n-\xi_{-\vec{k}+\vec{q}_1}}\frac{1}{k_f} (\vec{k}-\frac{\vec{q}_1}{2} )\cdot \vec{\sigma}i\sigma_2
\big]\nn\\
&=& -\int \frac{d^3\vec{k}}{(2\pi)^3} \frac{1}{\beta} \sum_{i\omega_n}
\frac{1}{k_f}(k_{\mu}-\frac{q_{1\mu}}{2}) \frac{(-i\omega_n+\xi_{\vec{k}}) (i\omega_n+\xi_{\vec{k}+\vec{q}_2}) (i\omega_n+\xi_{\vec{k}-\vec{q}_1-\vec{q}_3 }) (i\omega_n+\xi_{\vec{k}-\vec{q}_1}) }{(\omega_n^2+\xi_{\vec{k}}^2) (\omega_n^2+\xi_{\vec{k}+\vec{q}_2}^2) (\omega_n^2+\xi_{\vec{k}-\vec{q}_1-\vec{q}_3}^2) (\omega_n^2+\xi_{\vec{k}-\vec{q}_1}^2) }.
\label{eq:D_b}
\eea
Denote $\mathcal{D}^{b(0,1,2,3)}_{\mu}$ to be the $\vec{q}_1=\vec{q}_2=\vec{q}_3=0$ term, the linear in $\vec{q}_1$ term, the linear in $\vec{q}_2$ term, and the linear in $\vec{q}_3$ term of Eq. (\ref{eq:D_b}), respectively.
The evaluations are similar to $\mathcal{D}^a_{\mu}$.
The results are
\bea
\mathcal{D}_{\mu}^{b(0)}&=& 0,
\eea
\bea
\mathcal{D}_{\mu}^{b(1)}&=& -2 N_f\frac{q_{1\mu}}{k_f} \int d\epsilon \frac{1}{\beta} \sum_{i\omega_n} \frac{\epsilon^2(\epsilon^2-3\omega_n^2)}{(\omega_n^2+\epsilon^2)^4} +\frac{1}{2}N_f\frac{q_{1\mu}}{k_f} \int d\epsilon \frac{1}{\beta} \sum_{i\omega_n} \frac{\epsilon^2-\omega_n^2}{(\omega_n^2+\epsilon^2)^3},
\eea
\bea
\mathcal{D}_{\mu}^{b(2)}&=&N_f \frac{q_{2\mu}}{k_f}\int d\epsilon \frac{1}{\beta} \sum_{i\omega_n} \frac{\epsilon^2(\epsilon^2-3\omega_n^2)}{(\omega_n^2+\epsilon^2)^4},
\eea
\bea
\mathcal{D}_{\mu}^{b(3)}&=& -N_f\frac{q_{3\mu}}{k_f} \int d\epsilon \frac{1}{\beta} \sum_{i\omega_n} \frac{\epsilon^2(\epsilon^2-3\omega_n^2)}{(\omega_n^2+\epsilon^2)^4}.
\eea

Finally we consider the diagram Fig. \ref{fig:diagram} $(c)$.
The expression for $\mathcal{D}^c_{\mu}$ is
\bea
\mathcal{D}^b_{\mu}&=&\int \frac{d^3\vec{k}}{(2\pi)^3} \frac{1}{\beta} \sum_{i\omega_n}
\text{Tr}\big[
\frac{1}{-i\omega_n-\xi_{\vec{k}}} \frac{1}{-i\omega_n-\xi_{\vec{k}+\vec{q}_3}}i\sigma_2   \frac{1}{i\omega_n-\xi_{-\vec{k}-\vec{q}_2-\vec{q}_3}}\frac{1}{2}\sigma_{\mu}
\frac{1}{i\omega_n-\xi_{-\vec{k}+\vec{q}_1}}\frac{1}{k_f} (\vec{k}-\frac{\vec{q}_1}{2} )\cdot \vec{\sigma}i\sigma_2
\big]\nn\\
&=& -\frac{1}{k_f}\int \frac{d^3\vec{k}}{(2\pi)^3} \frac{1}{\beta} \sum_{i\omega_n}
\frac{1}{k_f}(k_{\mu}-\frac{q_{1\mu}}{2}) \frac{(-i\omega_n+\xi_{\vec{k}}) (-i\omega_n+\xi_{\vec{k}+\vec{q}_3}) (i\omega_n+\xi_{\vec{k}+\vec{q}_2+\vec{q}_3 }) (i\omega_n+\xi_{\vec{k}-\vec{q}_1}) }{(\omega_n^2+\xi_{\vec{k}}^2) (\omega_n^2+\xi_{\vec{k}+\vec{q}_3}^2) (\omega_n^2+\xi_{\vec{k}+\vec{q}_2+\vec{q}_3}^2) (\omega_n^2+\xi_{\vec{k}-\vec{q}_1}^2) }.
\label{eq:D_c}
\eea
Denote $\mathcal{D}^{c(0,1,2,3)}_{\mu}$ to be the $\vec{q}_1=\vec{q}_2=\vec{q}_3=0$ term, the linear in $\vec{q}_1$ term, the linear in $\vec{q}_2$ term, and the linear in $\vec{q}_3$ term of Eq. (\ref{eq:D_c}), respectively.
The results are
\bea
\mathcal{D}_{\mu}^{c(0)}&=& 0,
\eea
\bea
\mathcal{D}_{\mu}^{c(1)}&=& - N_f\frac{q_{1\mu}}{k_f} \int d\epsilon \frac{1}{\beta} \sum_{i\omega_n} \frac{\epsilon^2}{(\omega_n^2+\epsilon^2)^3} +\frac{1}{2}N_f\frac{q_{1\mu}}{k_f} \int d\epsilon \frac{1}{\beta} \sum_{i\omega_n} \frac{1}{(\omega_n^2+\epsilon^2)^2},
\eea
\bea
\mathcal{D}_{\mu}^{c(2)}&=&N_f \frac{q_{2\mu}}{k_f}\int d\epsilon \frac{1}{\beta} \sum_{i\omega_n} \frac{\epsilon^2}{(\omega_n^2+\epsilon^2)^3},
\eea
\bea
\mathcal{D}_{\mu}^{c(3)}&=& 2N_f\frac{q_{3\mu}}{k_f} \int d\epsilon \frac{1}{\beta} \sum_{i\omega_n} \frac{\epsilon^2}{(\omega_n^2+\epsilon^2)^3}.
\eea

Define $D^{(i)}$ ($i=1,2,3$) as the coefficient of $q_{i\mu}$ in $\mathcal{D}^{a(i)}_{\mu}+\mathcal{D}^{b(i)}_{\mu}+\mathcal{D}^{c(i)}_{\mu}$.
We obtain
\bea
D^{(1)}&=&N_f \frac{1}{k_f} \int d\epsilon \frac{1}{\beta} \sum_{i\omega_n} \frac{-\omega_n^4-5\epsilon^4+18\omega_n^2\epsilon^2}{2(\omega_n^2+\epsilon^2)^4},
\eea
\bea
D^{(2)}&=&N_f \frac{1}{k_f} \int d\epsilon \frac{1}{\beta} \sum_{i\omega_n} \frac{-4\omega_n^2\epsilon^2}{(\omega_n^2+\epsilon^2)^4},
\eea
\bea
D^{(3)}&=&2N_f \frac{1}{k_f} \int d\epsilon \frac{1}{\beta} \sum_{i\omega_n} \frac{\epsilon^2}{(\omega_n^2+\epsilon^2)^3}.
\eea
Using
\bea
\int d\epsilon \frac{1}{\beta} \sum_{i\omega_n} \frac{\epsilon^4}{(\omega_n^2+\epsilon^2)^4}= \int d\epsilon \frac{1}{\beta} \sum_{i\omega_n} \frac{\epsilon^2\omega_n^2}{(\omega_n^2+\epsilon^2)^4}=\frac{1}{5}\int d\epsilon \frac{1}{\beta} \sum_{i\omega_n} \frac{\omega_n^4}{(\omega_n^2+\epsilon^2)^4}
=\frac{7\zeta(3)}{8^3\pi^2}\frac{1}{T^2},
\eea
one arrives at
\bea
D^{(1)}=-D^{(2)}=D^{(3)}=\frac{1}{2}N_f\frac{1}{k_f}\frac{7\zeta(3)}{8^2\pi^2}\frac{1}{T^2}.
\eea
When the system is close to the superconducting transition point,
we can set $T=T_c$.
Each diagram in Fig. \ref{fig:diagram} represents eight terms in the Trlog-expansion of the free energy.
Since Fig. \ref{fig:diagram} is a fourth order term, there is an additional $\frac{1}{4}$ factor.
Combing these together, we arrive at the expression of $\Delta F^{(4)}$.

\section{Gapped surface states at interfaces}

\subsection{Gapped surface states at the interface of superconducting $p+is$ and $p-is$ bulks}
\label{sec:gapped_interface}

In this section, we solve for the surface states at the interface of superconducting $p+is$ and $p-is$ bulks.
We consider the surface $\Gamma$-point, i.e. $k_x=k_y=0$.
The surface spectrum has a gap equal to $\Delta_p$.

For convenience, in this section, we use a coordinate system, such that the interface lies in the $xy$ plane, which is different from what is chosen in Fig. 2 of the main text.
The spatial distribution of the singlet pairing component is taken as $\Delta_s(z)=\Delta_s$ at $z<0$ and $\Delta_s(z)=-\Delta_s$ at $z>0$,
where $\Delta_s$ is assumed to be positive.
At the surface $\Gamma$-point, the eigen-equation is
\bea
\big(\frac{\hbar^2}{2m} (-\partial_z^2-k_f^2)\tau_3+\frac{\Delta_p}{k_f} (-i\partial_z)\sigma_1\tau_1-\Delta_s(z) \sigma_2\tau_1   \big)\Psi(z)
=\epsilon \Psi(z),
\label{eq:eigen}
\eea
in which $\epsilon$ is the energy of the surface state.
The boundary conditions are $\Psi(z\rightarrow\pm\infty)\rightarrow 0$, $\Psi(z\rightarrow 0+)=\Psi(z\rightarrow 0-)$, and $\partial_z\Psi(z\rightarrow 0+)=\partial_z\Psi(z\rightarrow 0-)$.
Plugging the trial wavefunction $\Psi(z)=\Phi e^{ik_z z}$ into Eq. (\ref{eq:eigen}),
we obtain
\bea
\big(\frac{\hbar^2}{2m} (k_z^2-k_f^2)\tau_3+\frac{\Delta_p}{k_f} k_z\sigma_1\tau_1-\Delta_s(z) \sigma_2\tau_1-\epsilon\big)\Phi=0,
\label{eq:eigen2}
\eea
in which $+$ ($-$) is for $z<0$ ($z>0$).

Eq. (\ref{eq:eigen2}) can be factorized into two independent equations in the $S_z=\pm \frac{1}{2}$ sectors.
In the following, we consider the $S_z=\frac{1}{2}$ sector.
The $S_z=-\frac{1}{2}$ sector can be solved similarly.
In the weak pairing limit, i.e., $\Delta_s,\Delta_p\ll\frac{\hbar^2k_f^2}{2m}$,
the momentum $k_z$ can be approximated as $k_z=k_f(\eta - i\nu\xi)$,
in which $\nu=1$ when $z<0$ and $\nu=-1$ when $z>0$, and $\eta=\pm 1$.
$\xi$ is positive due to the boundary condition $\Psi(z\rightarrow\pm\infty)\rightarrow 0$.
For each $(\nu,\eta)$, there is a solution $\Phi_{\nu\eta}$, as
\bea
\Phi_{\nu,\eta}&=&\left( \begin{array}{c}
\eta\Delta_p+i\nu\Delta_s\\
i\nu\eta \sqrt{\Delta_p^2+\Delta_s^2-\epsilon^2}+\epsilon
\end{array}\right),
\eea
with the corresponding $\xi_{\nu\eta}$ given by $\xi_{\nu\eta}=\frac{\sqrt{\Delta_p^2+\Delta_s^2-\epsilon^2}}{2}$.
Then the wavefunction $\Psi(z)$ is given by
\bea
\Psi(z)&=&\sum_{\nu=\pm 1,\eta=\pm 1}  C_{\nu,\eta} \Phi^{\uparrow}_{\nu,\eta}e^{i\eta k_f z} e^{\nu\frac{\sqrt{\Delta_p^2+\Delta_s^2-\epsilon^2}}{2} z} \Theta(-\nu z),\label{eq:interfacewavef}
\eea
in which $\Theta$ is the step function defined by $\Theta(x)=1$ when $x>0$ and $\Theta(x)=0$ when $x<0$.

Plugging Eq. (\ref{eq:interfacewavef}) into the boundary conditions at $z=0$, i.e., $\Psi(z\rightarrow 0+)=\Psi(z\rightarrow 0-)$ and $\partial_z\Psi(z\rightarrow 0+)=\partial_z\Psi(z\rightarrow 0-)$, we obtain
\bea
\det\left( \begin{array}{cc}
\{ \Phi_{\nu,\eta} \} \\
\{ \eta\Phi_{\nu,\eta}\}
\end{array}\right)=0,
\label{eq:bdMat}
\eea
in which $\{ \Phi_{\nu,\eta} \}$ is the abbreviation of the $2\times 4$ matrix $\{\Phi^{\uparrow}_{+,+},\Phi^{\uparrow}_{+,-},\Phi^{\uparrow}_{-,+},\Phi^{\uparrow}_{-,-}\}$, and similar for $\{ \eta\Phi_{\nu,\eta} \}$.
The solution of Eq. (\ref{eq:bdMat}) is
\bea
\epsilon= \Delta_p,
\eea
with a two-fold degeneracy.
The wavefunctions are given by
\bea
\Psi^{\uparrow}_{\pm}(z)&=&\frac{1}{\sqrt{\mathcal{N}}}\left(\begin{array}{c}
1\\
0\\
0\\
\pm1
\end{array}\right)
e^{\pm ik_fz} e^{-\frac{\Delta_s}{2k_f}|z|}.
\eea

For the $S_z=-\frac{1}{2}$ sector,
there are two solutions with $\epsilon=-\Delta_p$, which are given by
\bea
\Psi^{\downarrow}_{\mp}(z)&=&\frac{1}{\sqrt{\mathcal{N}}}\left(\begin{array}{c}
0\\
\pm 1\\
1\\
0
\end{array}\right)
e^{\mp ik_fz} e^{-\frac{\Delta_s}{2k_f}|z|}.
\eea

\subsection{Gapped surface states at the interface of the superconducting $p+is$ and $-p+is$ bulks}
\label{sec:gapped_interface2}

In this section, we consider the surface states at the interface of superconducting $p+is$ and $-p+is$ bulks.
We show that the surface spectrum has a gap which is equal to $\Delta_s$.

The coordinate system is chosen such that the interface is within the $xy$-plane.
The singlet component is spatially uniform, and the triplet component  is taken as
$\Delta_p(z)=\Delta_p$ when $z<0$ and $\Delta_p(z)=-\Delta_p$ when $z>0$,
where $\Delta_p$ is assumed to be positive.
For simplicity, we consider the surface $\Gamma$-point.
The eigen-equation is
\bea
\big(\frac{\hbar^2}{2m} (-\partial_z^2-k_f^2)\tau_3+\frac{\Delta_p(z)}{k_f} (-i\partial_z)\sigma_1\tau_1-\Delta_s \sigma_2\tau_1   \big)\Psi(z)
=\epsilon \Psi(z),
\label{eq:delta_p}
\eea
with the boundary conditions $\Psi(z\rightarrow\pm\infty)\rightarrow 0$, $\Psi(z\rightarrow 0+)=\Psi(z\rightarrow 0-)$, and $\partial_z\Psi(z\rightarrow 0+)=\partial_z\Psi(z\rightarrow 0-)$.

Again Eq. (\ref{eq:delta_p}) can be factorized into the $S_z=\pm \frac{1}{2}$ factors.
The procedure for obtaining the surface states is exactly similar as in the former section.
Here we summarize the results.
The two wavefunctions with $\epsilon=-\Delta_s$ are
\bea
\Psi^{\uparrow}_{\pm}(z)&=& \frac{1}{\sqrt{\mathcal{N}}} \left( \begin{array}{c}
1\\
0\\
0\\
i
\end{array}\right)
e^{\pm ik_f z} e^{-\frac{\Delta_p}{2k_f}|z|},
\eea
and the two wavefunctions with $\epsilon=\Delta_s$ are
\bea
\Psi^{\downarrow}_{\mp}(z)&=& \frac{1}{\sqrt{\mathcal{N}}} \left( \begin{array}{c}
0\\
-i\\
1\\
0
\end{array}\right)
e^{\mp ik_f z} e^{-\frac{\Delta_p}{2k_f}|z|}.
\eea

\subsection{The chiral ``sheet" mode}

\begin{figure}
        \includegraphics[width=0.5\textwidth]{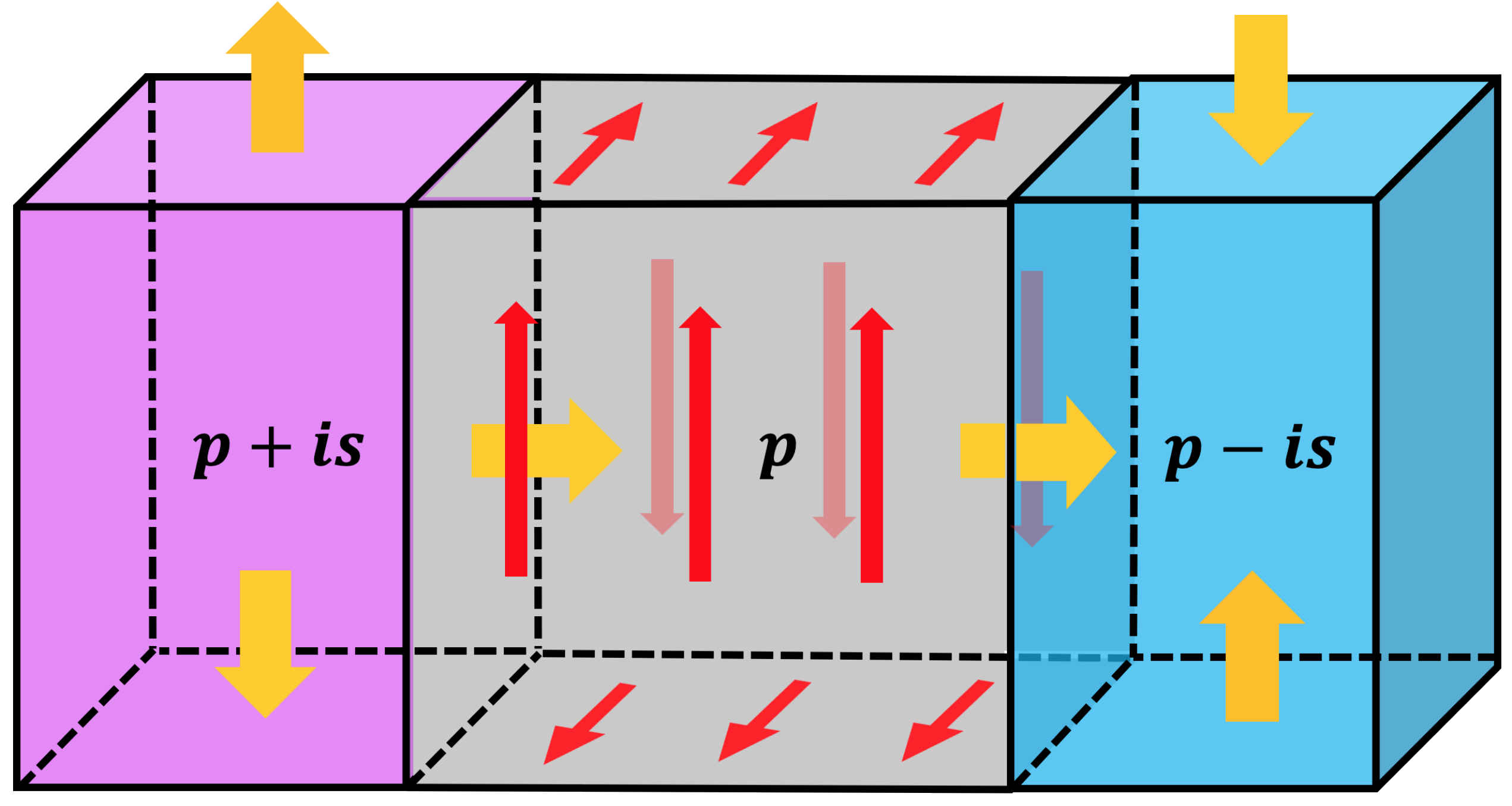}
\caption{A $p$-wave bulk sandwiched between the $p\pm is$ films.
The yellow arrows represent the directions of magnetizations on the surfaces or interfaces.
The red arrows show the propagation directions of the ``chiral sheet" mode on the upper and lower side-surfaces of the $p$-wave bullk.
}
\label{fig:p_sandwich}
\end{figure}

The chiral Majorana mode can also be viewed as propagating along the edge of the interface between the $p+is$ and the $p-is$ bulks.
To understand this, the domain interface where $\Delta_s$ changes sign is softened to a $p$-wave bulk sandwiched between the $p\pm is$ films
as shown in Fig. \ref{fig:p_sandwich}.
The gapless Majorana-Dirac fermions on the left and right side-surfaces of the $p$-wave bulk
acquire masses with opposite signs due to their proximity with the superconducting films of opposite time reversal symmetry breaking patterns.
The Majorana-Dirac fermion remains gapless on the other four side-surfaces of the $p$-wave bulk.
There exists circulating ``chiral sheet" mode \cite{Lee2009} on these four side-surfaces
 since they  form a fattened domain wall where the mass changes sign.


\end{widetext}

\end{document}